\documentclass[manuscript]{acmart}
\AtBeginDocument{%
  }

\setcopyright{acmlicensed}
\copyrightyear{2026}
\acmYear{2026}
\acmDOI{XXXXXXX.XXXXXXX}

\acmISBN{978-1-4503-XXXX-X/2026/06}
\newcommand{\system}[0]{GraphTide}
\usepackage[normalem]{ulem}
\usepackage{enumitem}
\usepackage{environ}
\usepackage{listings}
\acmSubmissionID{5319}

\begin{document}

\title{\system{}: Augmenting Knowledge-Intensive Text with Progressive Nested Graph}

\author{Xin Qian}
\email{qianxin@zju.edu.cn}
\affiliation{%
  \institution{State Key Lab of CAD\&CG, Zhejiang University}
  \city{Hangzhou}
  \state{Zhejiang}
  \country{China}
}

\author{Dazhen Deng}
\email{dengdazhen@zju.edu.cn}
\affiliation{%
  \institution{School of Software Technology, Zhejiang University}
  \city{Ningbo}
  \state{Zhejiang}
  \country{China}
}

\author{Zhaoping He}
\email{hezhahoping8@gmail.com}
\affiliation{%
  \institution{State Key Lab of CAD\&CG, Zhejiang University}
  \city{Hangzhou}
  \state{Zhejiang}
  \country{China}
}

\author{Xingbo Wang}
\email{xingbo.wang@us.bosch.com}
\affiliation{%
  \institution{Bosch Research North America \& Bosch Center for Artificial Intelligence (BCAI)}
  \city{Sunnyvale}
  \state{California}
  \country{United States}
}

\author{Yuchen He}
\email{heyuchen@zju.edu.cn}
\affiliation{%
 \institution{State Key Lab of CAD\&CG, Zhejiang University}
  \city{Hangzhou}
  \state{Zhejiang}
  \country{China}
}

\author{Yingcai Wu}
\email{ycwu@zju.edu.cn}
\affiliation{%
 \institution{State Key Lab of CAD\&CG, Zhejiang University}
  \city{Hangzhou}
  \state{Zhejiang}
  \country{China}
}

\renewcommand{\shortauthors}{Qian et al.}

\begin{abstract}
Knowledge-intensive text usually contains fruitful entities and complex relationships, such as academic articles and scientific exposition. Reading and comprehending such texts often demands considerable time and mental effort to track the relationships between entities.
To reduce the burden, we present \system{}, a visualization technique that progressively constructs nested entity-relationship graphs with animation to support the understanding of complex text. Our method features an on-demand entity-relationship decomposition pipeline that constructs nested graphs to represent intra- and inter-sentence relationships. Moreover, we propose a structure-aware force-directed layout optimization algorithm to enhance structural clarity. Sentences and their associated entities are incrementally revealed through animated transitions, helping users maintain context as the narrative unfolds.
A user study shows that \system{} significantly improves users' comprehension of knowledge-intensive texts compared to traditional graph-based techniques and static nested graph representations.
\end{abstract}



\begin{teaserfigure}
  \includegraphics[width=\textwidth]{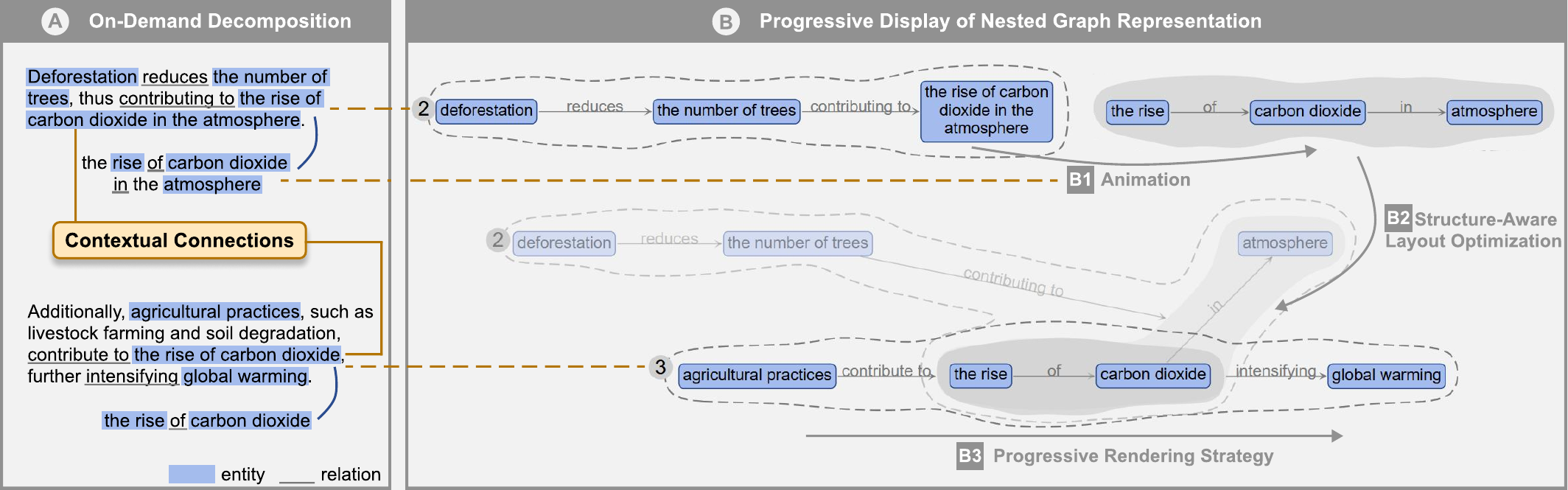}
  \caption{\system{} supports the progressive understanding of knowledge-intensive text by incrementally constructing and displaying nested entity-relationship graphs. (A) \system{} extracts entities and relationships from complex sentence structures while preserving contextual connections across sentences. (B) \system{} progressively displays nested graph structures through (B1) animation, (B2) structure-aware layout optimization that maintains spatial alignment with textual positions, and (B3) a progressive display strategy that unfolds entity-relationship graphs in sync with the reading flow.}
  \label{fig:teaser}
\end{teaserfigure}

\received{20 February 2007}
\received[revised]{12 March 2009}
\received[accepted]{5 June 2009}

\maketitle

\section{Introduction}

Text remains the most prevalent medium for conveying information in both routine and specialized contexts~\cite{ReviewOfTextVisualization}.
However, reading and comprehending text remains cognitively demanding tasks, as readers must follow the flow of ideas and track interrelated entities. For long texts, summarization and navigation techniques have been widely explored to provide high-level overviews or help locate relevant segments. In contrast, \textbf{medium-length} texts at the paragraph or section level typically require close reading ~\cite{kintsch1978toward,britton1991using}, especially for \textbf{knowledge-intensive texts}, such as academic articles and scientific exposition, because their meaning depends on dense terminology and intricate inter-entity relationships, which cannot be easily grasped through skimming or selective navigation. 

Such knowledge-intensive texts highlight a fundamental limitation of plain text. 
As a linear sequence of symbols, text often lacks the expressive capacity to convey complex conceptual and relational structures~\cite{Graphologue}, making it difficult to identify how entities relate to one another.
Furthermore, the absence of explicit contextual linking in verbose narratives places a heavy burden on readers' working memory~\cite{MemoryCapacity}, hindering their ability to construct a coherent mental model of the article's overall structure~\cite{EffectOfText}. These challenges call for \textbf{text augmentation} techniques that preserve the full content while assisting close reading, complementing summarization approaches that primarily serve high-level overviews.

Graph-based visualization has emerged as an effective and intuitive approach of text augmentation, especially by capturing relationships among entities in text~\cite{KNowNEt,KG4Vis,KGScope,CommonsenseVIS}.
Representing entities and their relationships as nodes and edges offers a structured and clear overview of textual content, which is particularly useful for general texts.
However, when applied to knowledge-intensive texts, this approach faces several challenges.
First, existing methods often extract entities and relations from text and represent them directly as flat nodes and edges in a \textbf{single-layer graph}~\cite{Graphologue, Sensecape}.
This flat structure faces a trade-off: it either captures only the top-level semantic relations while ignoring the internal composition of subordinate clauses, or it decomposes every sentence into highly fine-grained units, resulting in overly complex and fragmented graph structures, particularly in the case of long or nested sentences. 
Second, \textbf{static and monolithic graph} representations are often misaligned with the natural reading process of humans, which unfolds in a sequential and incremental manner.
Readers typically process information linearly, from left to right and top to bottom, gradually integrating meaning across clauses and sentences.
In contrast, presenting the entire graph structure at once can lead to cognitive overload and make it difficult to maintain continuity between the evolving text and its corresponding graph.

In this study, we introduce \textbf{\system{}}, a visualization technique for text augmentation that supports the progressive understanding of knowledge-intensive texts while preserving their original content. 
To address the limited capacity of existing graph representations in capturing complex semantics, we propose a \textbf{nested graph representation} that captures both the internal composition within individual sentences and the contextual continuity across multiple sentences. 
To avoid overloading users with overly detailed or fragmented information, we construct the graph in a progressive and selective manner.
Specifically, each sentence is first decomposed into coarse-grained semantic units based on its syntactic structure, such as subject-verb-object triples and subordinate clauses.
These semantic units are then refined into finer-grained entities and relations only when they are complex sentence structure or their internal components are referenced or reused in context.
This on-demand decomposition allows the graph to evolve gradually, preserving shared concepts across sentences while avoiding unnecessary complexity.

However, a graph representation will break the linear layout of the original sentences.
To make the comprehension naturally aligned with original sentence reading order, we propose a method to ensure the spatial and temporal synchronization of the graph and text.
First, we propose a \textbf{structure-aware layout optimization algorithm} to ensure spatial alignment between the nested graph and the original sentence layout.
Since natural language follows a top-to-bottom, left-to-right reading order, the entities and phrases in a sentence also exhibit an implicit spatial structure.
Our algorithm incorporates this spatial prior by using the textual positions of entities as soft constraints within a force-directed layout model.
This approach preserves the overall aesthetic quality of the graph while ensuring that the relative positions of nodes correspond closely to their locations in the original text.
Second, \system{} adopts a \textbf{progressive rendering strategy} to ensure temporal consistency between the graph and the reading flow of the text.
Rather than presenting the entire nested graph at once, which may overwhelm users with complex structures, we reveal graph elements incrementally, both sentence by sentence and node by node, in sync with the textual progression.
To further support users’ cognitive flow, we incorporate carefully designed animations that ensure smooth and continuous transitions, helping maintain narrative context and user engagement as the graph evolves.

To support information review and summarization after the progressive rendering phase, \system{} integrates a set of interactive features that enable deeper exploration.
These include context-aware, bidirectional linking between text and graph, allowing users to highlight relevant nodes, edges, and textual mentions by hovering over either representation.
In addition, a ranked list of frequently mentioned entities helps users quickly locate salient concepts and trace their associated contexts within the graph, enabling efficient, entity-driven navigation.

We evaluated \system{} on medium-length knowledge-intensive texts, a common context for close reading, by comparing it with traditional graph-based techniques and static nested graph representations. Our study shows that \system{} improves comprehension, clarifies structure, reduces cognitive load, and supports contextual integration and efficient review.
In summary, the contributions of this work are: 
\begin{itemize}[leftmargin=*] 
\item A \textbf{nested graph representation} for knowledge-intensive texts that preserves complex sentence structures and enhances contextual continuity. 
\item \textbf{\system{}}, a visualization technique for text augmentation that incorporates a structure-aware layout optimization algorithm and a progressive rendering strategy to naturally present nested graphs. 
\item A \textbf{user study} demonstrating the effectiveness and usability of \system{} in supporting the progressive comprehension of medium-length knowledge-intensive text. 
\end{itemize}

\section{Related Work}

In this section, we review three closely related research areas: text comprehension techniques, text visualization, and graph layout optimization.

\subsection{Text Comprehension Techniques}
To support text comprehension, prior work has explored a range of techniques designed to help users extract, organize, and make sense of textual information. These techniques can be broadly categorized into two types: summarization approaches that condense the original content for efficient access, and augmentation approaches that preserve the original text while enhancing user understanding.

Summarization approaches aim to reduce information redundancy and present high-level summaries for efficient access. Prior work includes user-driven tools that support flexible information organization including graph-based systems~\cite{MindManager,Sandbox,Sensecape,texSketch} and spatial interfaces~\cite{LiquidText}, as well as automated methods that extract graphs from text, either at the document level~\cite{Jigsaw,MultiDocument_Graph} or the entity level~\cite{EntityWorkspace,Storytelling,NetLens,SentinelVisualizer}. While effective in surfacing key content, such approaches often omit critical details and nuanced relationships, resulting in incomplete representations that hinder thorough comprehension of the original text.

In contrast, augmentation approaches aim to preserve the original text while enhancing user comprehension through visual or structural emphasis. For example, Graphologue~\cite{Graphologue} transforms language model responses into interactive node-link diagrams, enabling users to explore and reorganize information extracted from long-form responses in a non-linear and graphical manner. GP-TSM~\cite{GP-TSM} modifies the visual presentation of text itself by gradually fading less central content, making key information more salient without altering the grammar or omitting details. These methods maintain the full textual content and offer alternative interaction mechanisms to support comprehension. However, these methods offer limited support for capturing nested semantic structures and sustaining contextual continuity across sentences, which are essential for understanding knowledge-intensive texts.

Therefore, we propose a new approach that preserves the full textual content while progressively revealing nested entity-relationship structures to support the comprehension of knowledge-intensive texts. Through nested graph representation as well as spatial and temporal alignment with the natural reading flow, this approach addresses the limitations of prior methods in conveying complex semantics and maintaining contextual continuity.

\subsection{Text Visualization}

Text visualization~\cite{SoS_TextVis} is a technique to visually present textual information, aiming to expose the structural or semantic characteristics of text for easier exploration and analysis. Early techniques focus on typographic visualizations, including text highlighting~\cite{TextHighlight1,TextHighlight2,TextHighlight3} and word clouds~\cite{WordCloud1,WordCloud2,WordCloud3,WordCloud4}, which emphasize lexical salience. To support deeper analysis and exploration, structured visualizations such as charts~\cite{Chart1,Chart2,Chart3,Chart4}, timelines~\cite{Timeline1,Timeline2}, graphs~\cite{Graph1,Graph2,Graph3}, and tree-based layouts~\cite{PhraseNet,WordTree,SentenTree} have been introduced to reveal various semantic patterns and patterns within the text.

However, existing text visualization techniques remain limited in supporting the comprehension of knowledge-intensive texts. Typographic approaches (e.g., highlights, word clouds) focus on surface-level lexical features while lacking semantic depth. Structured methods such as charts, timelines, and tree-based layouts emphasize frequency, temporal flow, or local syntactic patterns, yet are less effective in capturing entity-level semantics or expressing logical relations among entities.

Graph-based techniques offer greater expressive power by explicitly modeling entities and their relationships. However, most existing methods represent these relations as flat, single-layer graphs that either oversimplify nested sentence structures or fragment them into overly fine-grained units. Moreover, such static and monolithic representations often fail to align with the sequential and incremental nature of human reading, making it difficult for users to maintain contextual continuity and gradually integrate meaning across complex texts. To address these challenges, we propose a technique that progressively constructs nested entity-relationship graphs that effectively capture complex semantics and maintain contextual continuity.

\subsection{Graph Layout Optimization}
In graph visualization, layout optimization plays a crucial role in enhancing both readability and the effectiveness of information representation. The most basic form of a node-link diagram is the flat graph, in which there are no inherent hierarchical relationships among nodes. Common layout strategies for flat graphs include geometric layout, hierarchical layout~\cite{Dagre,Sugiyama} and force-directed layout~\cite{ForceLayout1,ForceLayout2,ForceLayout3}. Geometric layouts arrange nodes based on predefined spatial patterns, such as grid, circular, or radial placements. Hierarchical layouts organize nodes according to directional or dependency structures, while force-directed layouts simulate physical systems to produce visually balanced configurations. 

In many practical scenarios, graph structures exhibit nested topology, where nodes can recursively contain other nodes. These are known as compound graphs~\cite{CompoundGraph_Definition}, characterized by both hierarchical containment and inter-level connections. Due to their structural complexity, layout algorithms must account for nested regions and edges across different levels. Existing automatic layout techniques for compound graphs can be broadly categorized into layered approaches ~\cite{CompoundGraph_Layer_Layout1,CompoundGraph_Layer_Layout2,CompoundGraph_Layer_Layout3} and force-based approaches ~\cite{CompoundGraph_Force_Layout1,CompoundGraph_Force_Layout2}. Layered approaches divide the drawing space into layers, sorting nodes for a consistent edge direction, typically from top to bottom, while force-based methods extend traditional force-directed models to handle nested groups and cross-level edges.

Existing compound graph layout methods assume a strictly nested structure without overlap between regions. However, nested graphs derived from knowledge-intensive texts often involve overlapping substructures, where different high-level entities share common subnodes due to sentence composition and contextual reuse. Such overlapping structures are not supported by standard compound graph models~\cite{CompoundGraph_Definition}. To address this, we propose a nested graph representation that supports flexible containment and cross-level connections, along with a structure-aware layout algorithm that integrates textual structure into the layout process. This ensures spatial alignment between graph elements and their corresponding positions in the text, enhancing readability and semantic continuity.
\section{Formative Study}
\label{sec:formative_study}

We conducted a formative study to examine how users read and comprehend texts in their everyday practices, with particular attention to the typical scales and structures of texts, the challenges users encounter, and the limitations of current tools. To clarify the textual scope for comprehension support and to inform the design objectives of \system{}, the study sought to answer the following questions:

\begin{itemize}
\item \textbf{Q1}: In which contexts do users engage with texts for reading and comprehension?
\item \textbf{Q2}: What challenges do users encounter in reading and comprehension?
\item \textbf{Q3}: What limitations exist in the tools that users rely on for reading assistance?
\end{itemize}


\subsection{Study Design}

We recruited ten participants (P1---P10) with varied educational levels, including undergraduates and postgraduates. Their academic backgrounds covered a range of disciplines such as computer science, information science, and education. 
All participants reported regular engagement with academic texts in their studies or work, ensuring that they had substantial experience with in-depth reading and comprehension in everyday practices.  

The study was structured around the three guiding questions \textbf{(Q1---Q3)}. In the first phase, participants were asked to recall and describe their everyday reading practices, with attention to both the kinds of texts they most frequently encounter and the ones they find particularly challenging to comprehend \textbf{(Q1)}. 
To provide a common ground for comparison across participants, they were also asked to read several passages of Wikipedia and summarize its main points, which provided a concrete basis for eliciting and discussing specific comprehension challenges \textbf{(Q2)}.  


In the second phase, participants were introduced to several visualization tools. Building on their prior experiences with reading aids and reflections on these state-of-the-art tools, participants were asked to identify the limitations of existing techniques for supporting comprehension \textbf{(Q3)}.

\subsection{In which contexts do users engage with texts for reading and comprehension?}

To address \textbf{Q1}, we asked participants to recall and describe their everyday reading practices, with a focus on both the types and scales of texts they frequently encountered and those they found particularly challenging to comprehend. For the purpose of the study, we distinguished between two broad types of texts: knowledge-intensive materials, defined as fact-rich texts that require deep integration and coherence tracking (e.g., academic articles, policy documents, or expository writing)~\cite{Rapid}, and less information-dense materials, which primarily serve expressive or narrative purposes rather than the communication of factual knowledge (e.g., essays or literary prose). 
We further categorized scale by textual granularity: short texts at the sentence or snippet level, medium-length texts at the paragraph or section level (typically 100–300 words, often treated as the basic unit of comprehension~\cite{kintsch1978toward,britton1991using}), and long texts spanning entire articles, reports, or books. This categorization offered a common frame of reference for eliciting participants' accounts of their reading contexts and comprehension practices.

Participants reported engaging with both knowledge-intensive and less information-dense texts in their everyday practices. While both categories were part of their regular reading, all noted that knowledge-intensive texts demanded substantially more effort to comprehend. As P5 explained, \textit{``Policy documents are not very long, but they are dense and require me to slow down.''} Similarly, P9 remarked, \textit{``When I need to learn something new, I usually turn to papers or technical notes, and that always takes much more concentration than reading essays or opinion pieces.''} By contrast, less information-dense texts such as essays were described as easier to \textit{``skim for gist''}(P6) and to \textit{``read more quickly''}(P1).

With respect to scale, participants reported working across short, medium, and long texts, but described different ways of engaging depending on length. Short texts such as headlines or snippets were often \textit{``read at a glance''}(P2) to capture surface-level information. For long texts, eight out of ten participants reported that they would locate relevant portions and focus attention there, since not all sections were equally valuable for their purposes. As P3 explained, \textit{``I usually focus on a section that is relevant to my question, going through the whole paper would be too much at once.''} In contrast, medium-length texts at the paragraph or section level were most frequently described as the unit at which participants engaged in sustained reading and comprehension, often requiring them to \textit{``slow down''}(P5) and to \textit{``concentrate on details''}(P9).

Taken together, participants' accounts suggest that medium-length, knowledge-intensive texts constitute a particularly salient domain of reading. Such texts were reported as both commonly encountered in study and work contexts and consistently described as demanding sustained effort to comprehend. This combination of prevalence and difficulty clarifies the textual scope for comprehension support, pointing to medium-length, knowledge-intensive passages as the critical target for \system{}.

\subsection{What challenges do users encounter in reading and comprehension?}

To examine \textbf{Q2}, we asked participants to reflect on difficulties encountered in reading medium-length, knowledge-intensive texts. We employed two complementary approaches: \textbf{semi-structured interviews}, in which participants recalled recent reading scenarios, and \textbf{a controlled reading task}, in which each participant read two medium-length passages from Wikipedia, one on cumulative effects and another on general relativity, and produced brief summaries. Across both activities, participants reported recurring difficulties in close engagement with such texts. We summarized three main challenges (C1---C3) as follows:

\begin{itemize}[leftmargin=*]
\item \textbf{C1. Mental Models are Hard to Maintain During Close Reading.}
Participants frequently reported that the mental models they constructed while reading were difficult to maintain, especially in dense texts rich with interrelated entities and concepts. Several noted that \textit{``I lose track of what concept relates to what''} (P4) and that \textit{``something just slips away''} (P6) as new information accumulated. They emphasized that linear text presentation \textit{``keeps going forward''} (P5) while what they needed was \textit{``a map of how these concepts connect''} (P9). In high-density passages, this mismatch left them struggling to keep entities and their interactions coherent.

\item \textbf{C2. Frequent Backtracking Disrupts Reading Flow.}
To make sense of current sentences, participants often needed to revisit earlier parts to recall prior definitions, references, or descriptions. They described \textit{``rereading the start of the section to remind myself what that term meant''} (P7) or \textit{``going back to check the setup of a causal chain''} (P3). Several participants commented that this effort was \textit{``exhausting, like doing double the work just to keep up''} (P2), and others noted that it \textit{``made me lose the thread of what I was reading''} (P6) whenever they looked back. Without support for quickly locating earlier content, participants were forced into repeated backtracking, which not only increased their cognitive effort but also disrupted their reading flow.

\item \textbf{C3. Key Entities and Relations Are Hard to Recall and Consolidate After Reading.}
When reflecting after reading, participants struggled to recall important entities and relations. Several admitted that they \textit{``just remember bits and pieces''} (P1), \textit{``retain fragments''} (P8), and \textit{``know I read the important part somewhere, but can't pull it out later''} (P5). Beyond simple recall, they found it difficult to consolidate what they had read into a coherent structure, as one participant explained that \textit{``I have pieces, but they don't fit together into something I can use later''} (P6). These fragmented recollections made it hard to build a reusable understanding, leaving participants uncertain about whether they had captured the essential content.

\end{itemize}

\subsection{What limitations exist in the tools that users rely on for reading assistance?}

To address \textbf{Q3}, we asked participants to reflect on the tools and strategies they typically use to aid comprehension and to evaluate their usefulness and limitations. We then introduced them to several state-of-the-art visualization tools for representing entity–relationship structures in text, such as Graphologue~\cite{Graphologue}, DEER~\cite{DEER}, and the Lokahi Prototype~\cite{LokahiPrototype}, and gathered their feedback on the effectiveness of these tools in supporting comprehension.


Several participants described relying on manual methods such as note-taking or sketching mind maps. While these strategies were considered \textit{``helpful for organizing thoughts''} (P1), they were also seen as demanding. As P6 explained, \textit{``Taking notes helps me clarify the main points, but it takes so much extra effort while I'm still trying to understand the text.''} Similarly, P3 noted, \textit{``With a mind map, I spend more time deciding how to draw the branches than actually reading.''}

Another commonly mentioned approach was the use of LLM-based tools to generate explanations or summaries. Participants voiced concerns about both accuracy and trust. As P2 put it, \textit{``Sometimes the explanation looks convincing, but I can't tell if it's really grounded in the text.''} P8 added, \textit{``Even if I ask it to quote the original sentences, I still need to go back and Ctrl+F to check everything myself.''} Several also noted that the outputs were often \textit{``too long''} (P4) and \textit{``full of extra commentary''} (P7), and as P3 summarized, \textit{``instead of helping me understand, it just makes me more distracted.''}

When introduced to state-of-the-art visualization tools, participants appreciated the idea of visualizing relationships between entities through graphical representations, but pointed out two main limitations. First, the single-layer representation was seen as \textit{``either too coarse or too fragmented''} (P8). As P3 remarked, \textit{``Either it just shows the big relations and skips the details, or it breaks the sentence into so many little pieces that I lose track of what it means.''} Second, static graph displays were misaligned with the natural reading process. As P7 explained, \textit{``I read line by line, building up the meaning gradually. But the graph just pops out all at once, and I can't follow the flow.''} These comments suggest that while visual tools hold potential, existing designs can impose cognitive load and disrupt reading rhythm.

In summary, participants reported that while existing tools provide partial support for comprehension, each also has notable drawbacks. Manual strategies demand substantial effort, LLM-based approaches raise concerns about accuracy and trust, and current visualization tools suffer from both structural limitations and temporal misalignment with natural reading practices. These limitations underscore the necessity of designing text augmentation techniques that align with natural reading processes while retaining the richness of the original content.
\section{Design Goals}



Building on the findings of the formative study (\autoref{sec:formative_study}), we identified medium-length, knowledge-intensive texts as both prevalent in practice and particularly challenging to comprehend, and thus as the primary focus of \system{}. The study also highlighted key comprehension challenges and limitations of existing reading assistance tools. In response, we propose four \textbf{D}esign Goals that guide the development of \system{} to enhance comprehension efficiency. We first introduce these design goals, and then use a scenario (\autoref{sec:usage_scenario}) to ground how these goals can be supported by \system{}.

\begin{itemize}[leftmargin=*]
\item \textbf{D1. Simplified Information Representation of Sentence Structure.}
A major challenge in understanding knowledge-intensive texts is that readers' mental models are difficult to maintain during close reading (\textbf{C1}). 
Existing approaches have attempted to address this by explicitly structuring entities and their relationships, for example through graph-based visualizations or hierarchical canvas designs. 
However, single-layer graphs, such as those used in Graphologue~\cite{Graphologue}, fall short in representing the nested structure of complex sentences, leading to representations that are either too coarse or too fragmented.
Similarly, hierarchical canvas-based designs, such as Sensecape~\cite{Sensecape}, organize information across multiple levels of abstraction but require users to switch between canvases, introducing interaction overhead and increasing cognitive load.
For knowledge-intensive text comprehension, we aim for a simplified visualization that explicitly reveals intra-sentence structural relationships and is intended to better align with readers' mental models, thereby supporting them in analyzing sentence structure and extracting key information more effectively.

\item \textbf{D2. Progressive Representation to Align with Mental Model Construction.}  
Presenting all information at once can lead to cognitive overload and make it difficult for readers to construct their mental models (\textbf{C1}).  
Prior work on progressive disclosure has shown that incrementally revealing information can reduce cognitive load and help users navigate complex content more effectively~\cite{progressive_disclousure}.  
Building on this insight, progressive representation for text comprehension aims to align with the narrative flow of the text by presenting information step by step.  
Instead of being overwhelmed by a fully expanded structure, readers can focus on one conceptual unit at a time, reducing mental strain and supporting the gradual construction of mental models.  
Moreover, by revealing information in the same sequence as the original text, progressive representation is intended to reduce the need for disruptive backtracking (\textbf{C2}) by keeping the current context visible and preserving continuity across successive units of information.

\item \textbf{D3. Convenient Representation of Contextual Relationships.}  
Prior work has shown that establishing contextual connections across sentences is essential for deep comprehension and coherence building in knowledge-intensive texts~\cite{context1,context2}.  
When new information is introduced, readers must relate it to previously mentioned entities to maintain coherence and construct a comprehensive mental model.  
However, because such connections are often dispersed across sentences, the absence of explicit guidance can force frequent backtracking, which disrupts the flow of reading (\textbf{C2}).  
To address this challenge, contextual relationships should be represented in a convenient and accessible manner, making it easier for readers to see how new content relates to prior information and to integrate it seamlessly into their evolving mental model.

\item \textbf{D4. Seamless Interaction Between Visual Representation and Text.}  
In knowledge-intensive text comprehension, structured visualizations can provide an effective means of representing complex relationships, but they should not be isolated from the original text.  
Readers often need to return to the text for details that extend beyond what the visualization conveys.  
To address this, we propose an interactive bidirectional link to connect the visual representation and the original text, so that elements in the visualization can be easily traced to their textual occurrences and vice versa.  
Such seamless navigation is aimed at enabling readers to leverage the combined strengths of both representations, thereby facilitating more efficient comprehension~\cite{DeFT}.

\item \textbf{D5. Entity-Driven Exploration for Review and Summarization.}  
Prior work has shown that entities provide effective anchors for extracting and summarizing information, as they often represent the central elements around which knowledge is organized~\cite{EntityWorkspace,LiquidText}.  
However, in knowledge-intensive texts, key entities and their relations are often difficult to recall and consolidate after reading (\textbf{C3}).  
To address this challenge, entity-driven exploration should be supported, enabling readers to trace important entities to their corresponding visual elements and textual contexts.  
Such a representation is intended to support efficient identification and summarization of core relationships, helping readers organize key information within the overall structure.

\end{itemize}

\section{Usage Scenario}
\label{sec:usage_scenario}
In this section, we describe a scenario that illustrates the workflow of \system{}, designed to support the above design goals for medium-length knowledge-intensive text comprehension. 

\begin{figure}[t]
  \centering
  \includegraphics[width=0.6\linewidth]{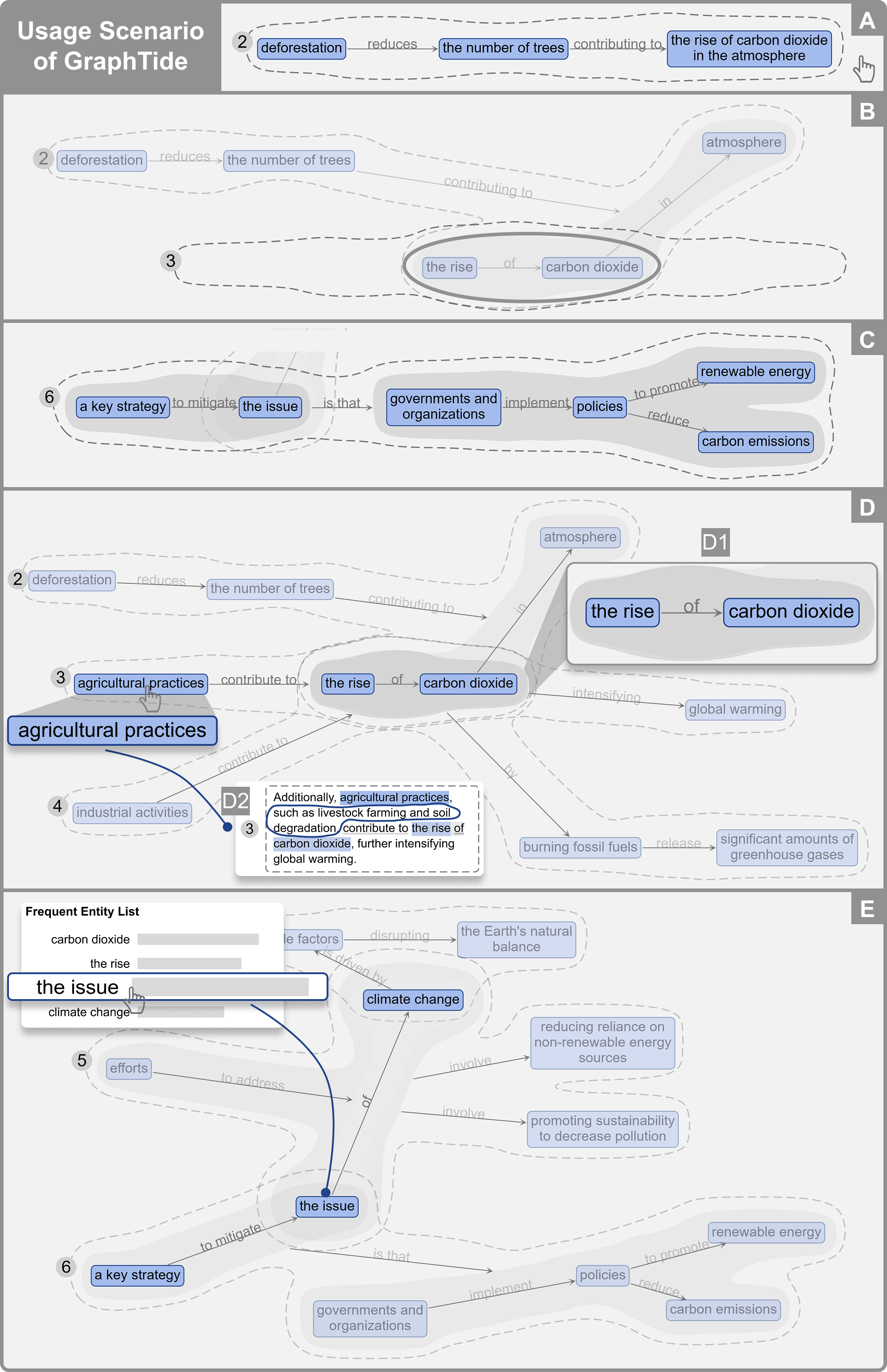}
  \centering
  \caption{Usage Scenario. Alex begins with a blank canvas, where each click triggers the progressive rendering of the next sentence (A). One node is decomposed and smoothly moved to the next sentence's region, signaling the continuation of the topic (B). The simplified structure helps Alex stay engaged (C). After reading, he quickly locates the ``carbon dioxide'' section of the graph (D1) and identifies three contributing factors. Alex hovers over ``agricultural practices'' to refer back to the original text for full details (D2). Finally, he clicks on frequent entity ``issue'' and identifies three solutions to climate change (E).}
  \Description{}
  \label{fig:scenario}
\end{figure}

Alex, a graduate student preparing for a climate policy exam, is reading a dense passage about human activities contributing to rising carbon dioxide levels.
He begins with a blank canvas, where each click progressively renders the nested graph of the next sentence \textbf{(D2)}, revealing entities and their relationships according to the original reading order (\autoref{fig:scenario}.A).
If the entire graph were presented at once, Alex might be overwhelmed by its complexity; progressive rendering instead keeps his reading flow aligned with the text.

As the visualization progresses, before the third sentence appears, he notices that the node ``the rise of carbon dioxide in the atmosphere'' is decomposed into finer-grained components: ``rise'' --- (``of'') --- ``carbon dioxide'' --- (``in'') --- ``atmosphere''. 
Moreover, the subgraph ``rise'' --- (``of'') --- ``carbon dioxide'' is smoothly moved downward into the region where the next sentence will appear (\autoref{fig:scenario}.B).
Unlike flat graphs that either oversimplify or fragment such structures, this decomposition preserves clarity while keeping related concepts together.
Alex thus develops a clear expectation that the upcoming sentence will continue discussing the rise of carbon dioxide \textbf{(D3)}.
During the reading process, he also finds that the nested entity-relationship graph presentation effectively reflects the structure of each sentence \textbf{(D1)}, keeping him highly engaged throughout comprehension (\autoref{fig:scenario}.C).

By this stage, Alex has already understood that the passage discusses several human activities contributing to the rise of carbon dioxide and their respective roles.
To revisit specific details, he refers back to the graph.
Relying on his memory of the visual structure formed during reading, he quickly locates the relevant subgraph centered on “carbon dioxide” (\autoref{fig:scenario}.D1).
By glancing at the graph, he immediately identifies the major contributing factors, namely deforestation, agricultural practices, and industrial activities.
Among these, Alex is less familiar with agricultural practices and seeks more detail.
He hovers over the corresponding node to return to the original text \textbf{(D4)}, where he finds examples such as livestock farming and soil degradation (\autoref{fig:scenario}.D2).
This helps him quickly clarify the meaning and deepen his understanding.

Beyond reviewing entities he already focused on, Alex also explores the frequent entity list.
He notices that “issue” is a frequent entity and clicks on it to locate its node in the graph \textbf{(D5)}.
With static representations, such connections would require scanning the entire graph, but here they are surfaced directly.
He sees that it appears in the final two sentences (\autoref{fig:scenario}.E), where he quickly identifies three proposed solutions: reducing reliance on non-renewable energy, promoting sustainability, and implementing policies.

Together, these interactions demonstrate how GraphTide addresses key comprehension challenges: preserving sentence structure, highlighting contextual continuity, and supporting entity-driven review.
\section{Methods}

As shown in \autoref{fig:teaser}, \system{} supports the progressive understanding of knowledge-intensive text through three complementary components. 
For each sentence, it performs on-demand entity–relationship decomposition (\autoref{fig:teaser}.A) to construct a \textbf{nested graph representation} that simplifies intra-sentence structure \textbf{(D1)} while preserving links to previously introduced content \textbf{(D3)}. Building on this, \system{} applies \textbf{structure-aware layout optimization algorithm} (\autoref{fig:teaser}.B2) to compute node positions to align with textual positions. When presented to users, \system{} employs a \textbf{progressive rendering strategy} (\autoref{fig:teaser}.B3) with animation (\autoref{fig:teaser}.B1) that keeps the display in sync with the reading flow to support gradual construction of mental models \textbf{(D2)}.

\subsection{Nested Graph Representation}

\begin{figure}[t]
  \centering
  \includegraphics[width=0.6\linewidth]{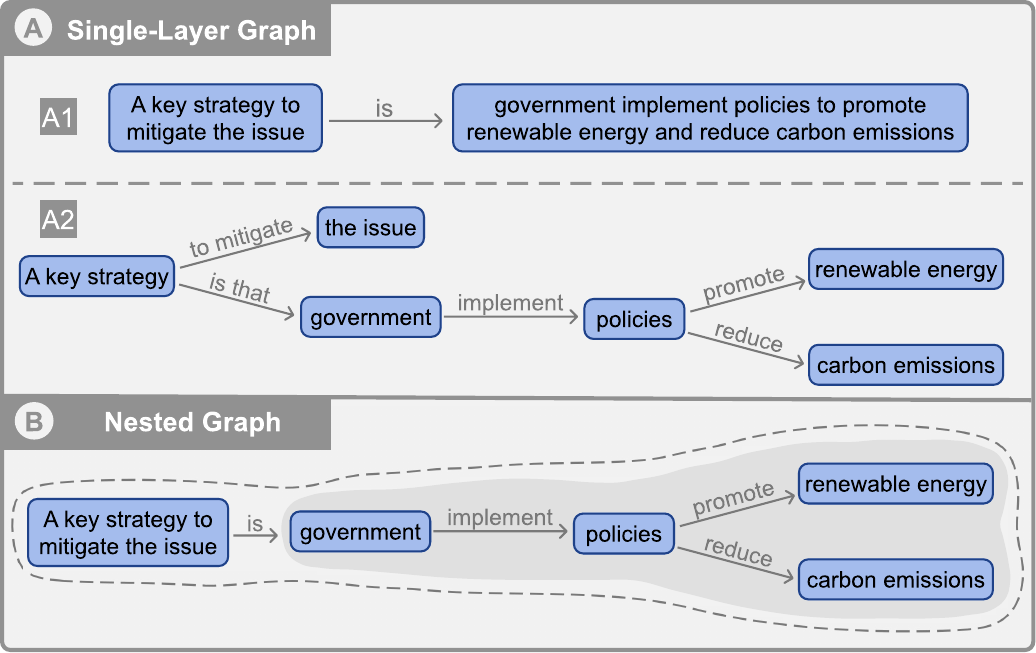}
  \centering
  \caption{Comparison of Single-Layer and Nested Graph Representations. A single-layer graph either captures only the top-level semantic relations (A1) or decomposes into highly fine-grained units (A2), both of which obscure semantic structure.In contrast, the nested graph (B) explicitly represents the nested entity–relationship structures, providing greater structural clarity.}
  \label{fig:flat_vs_nested}
\end{figure}

Existing approaches typically represent entities and relations in a \textbf{single-layer graph}~\cite{Graphologue,DEER,LokahiPrototype}. However, such flat structures struggle to capture the semantic composition of complex sentences. As illustrated in \autoref{fig:flat_vs_nested}.A, they captures only the top-level semantic relations that overlook internal composition (A1), or fragment sentences into overly fine-grained units that obscure key conceptual connections and increase cognitive load (A2). 
To address these limitations, \system{} introduces an \textbf{on-demand decomposition} strategy that incrementally refines semantic units into nested entity–relationship structures only when needed. This selective refinement avoids both oversimplification and excessive fragmentation, while preserving essential conceptual connections across sentences.  

Building on this, we further propose a \textbf{nested graph representation} (\autoref{fig:flat_vs_nested}.B) that makes the nested entity–relationship structures explicit and supports readers in capturing sentence-level composition together with cross-sentence continuity.

\subsubsection{On-Demand Decomposition} 
To systematically extract structured information from knowledge-intensive text, we employ an on-demand entity-relation decomposition. 
As shown in \autoref{fig:decomposition}.A, atomic decomposition extracts entity–relation triples from the input text and represents them as a single-layer node–link graph. At this stage, GPT-4o~\cite{GPT-4o} is instructed to annotate only the top-level entity–relation structure without further decomposition. To improve the quality of annotation, we detect two common errors: entities annotated without any associated relationships and relations involving non-existent entities. We then prompt GPT-4o to self-correct these inconsistencies.

Building on this, the decomposition pipeline progressively refines semantic units, as shown in \autoref{fig:decomposition}.B. The process starts with the extraction of the top-level entity–relation triples in the given sentence(B1). Entities that still encapsulate complex sentence structures can then be further decomposed (B2) to avoid overly coarse nodes and capture sentence structure with greater clarity. When new content introduces sub-concepts that overlap with previously mentioned entities, both the new entities (B3) and their corresponding earlier entities (B4) are decomposed to make the contextual connections explicit.

To assess the reliability of entity–relationship extraction, we conducted a small-scale evaluation on 16 passages of approximately 200 words each, drawn from diverse topics and domains. Eight passages were sampled from Wikipedia and eight were generated by a large language model, providing coverage of both encyclopedic and model-produced text. Each passage was processed with the on-demand decomposition pipeline, and all extracted entities and relationships were manually verified for correctness. As summarized in \autoref{tab:extraction}, GPT-4o achieved reasonably accurate performance for both entity and relation extraction, sufficient to support the construction of our nested graphs.

\begin{table}[t]
\centering
\caption{Performance of entity and relationship extraction.}
\begin{tabular}{ccc}
\toprule
\textbf{Metric} & \textbf{Entity Extraction} & \textbf{Relation Extraction} \\
\midrule
Total Items        & 1046 & 637 \\
Total Extracted Items    & 1069 & 601 \\
Correct Extracted Items      & 995 & 534 \\
\midrule
Precision (\%)      & 93.1\% & 88.9\% \\
Recall (\%)         & 95.1\% & 83.8\% \\
F1 Score (\%)       & 94.1\% & 86.3\% \\
\bottomrule
\end{tabular}
\label{tab:extraction}
\end{table}

\subsubsection{Visual Encoding of Nested Graphs}

\begin{figure}[t]
  \centering
  \includegraphics[width=0.6\linewidth]{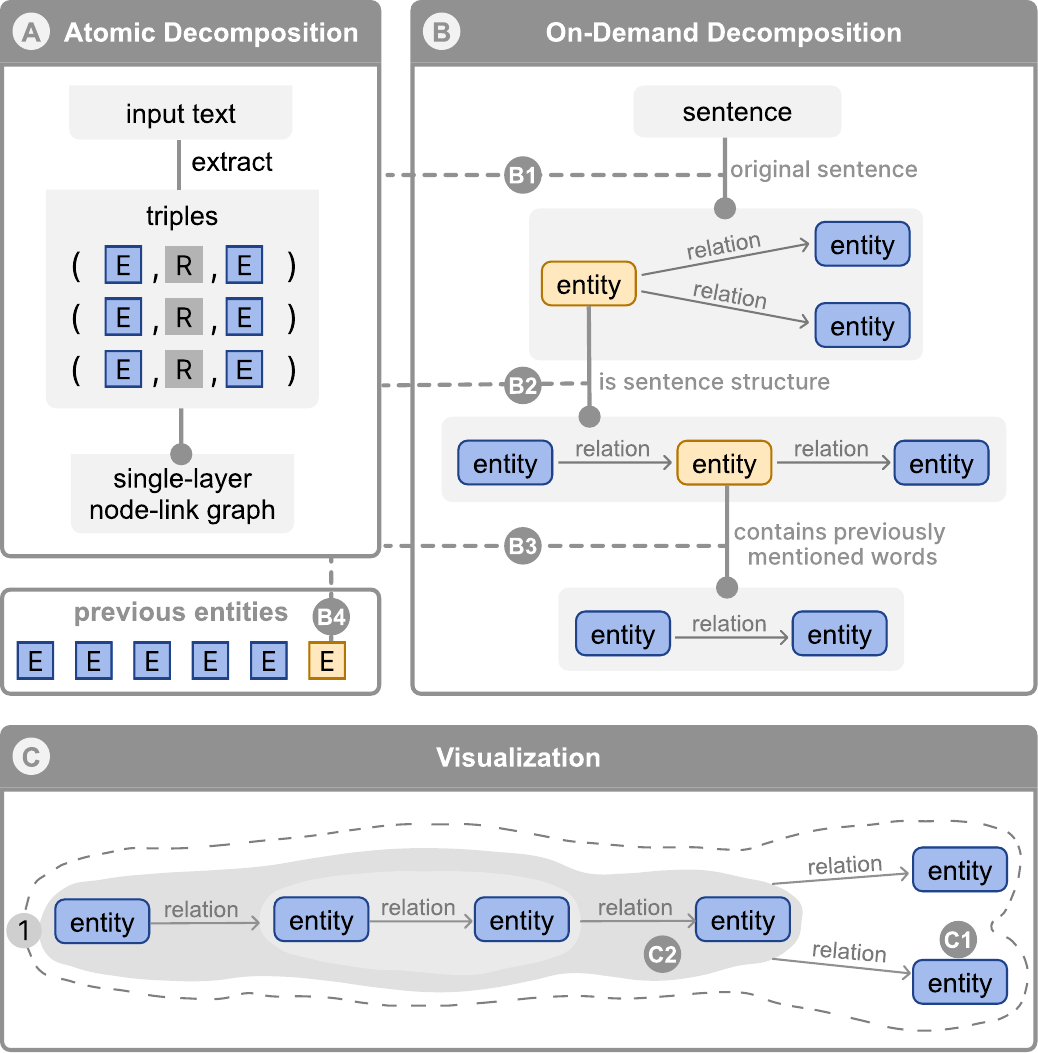}
  \centering
  \caption{Sentence Decomposition Pipeline and Corresponding Visualization: The decomposition pipeline performs multiple rounds of atomic decomposition (A), breaking down sentences into nested components (B), with the results visualized as a nested graph (C).}
  \label{fig:decomposition}
\end{figure}

Since the on-demand entity–relationship decomposition yields nested structures,  single-layer graphs are insufficient to represent the resulting composition. We therefore visualize the output as a nested graph (\autoref{fig:decomposition}.C). Each extracted entity is represented as a node, and nodes fall into two types depending on whether they are further decomposed. \textbf{Atomic nodes} (C1) remain non-decomposed and are drawn as labeled rectangles. \textbf{Composite nodes} (C2) encapsulate a subgraph of internal entities and relationships and are drawn as gray enclosing containers that contain their internal nodes and edges. Directed, labeled links encode relationships between entities. This visual encoding makes nested decomposition structure explicit while keeping the display compact, enabling users to quickly grasp the structure of a sentence and efficiently extract key information for better comprehension.

\subsection{Structure-Aware Layout Optimization}
\label{sec:layout}

In knowledge-intensive texts, sentences are visualized as nested graphs, where overlapping concepts across sentences create complex interconnections and impose additional spatial constraints. 
Furthermore, when the graph layout does not align with the top-to-bottom, left-to-right reading order of natural language, it can lead to confusion and disrupt the reader's cognitive flow. 
To effectively handle these cross-sentence overlaps and ensure alignment with the original sentence reading order, a more advanced layout optimization approach is required.


\subsubsection{Initial Layout Strategy} 
In our approach, the layout of each nested level is recursively determined using a consistent strategy. 
Within a single nested level, the connections between entities are generally not overly complex and can typically be represented as a directed acyclic graph. 
In such cases, we adopt a layered layout to preserve the logical structure of relationships and align with the left-to-right reading order of natural language. 
In rare instances where cycles exist, we first identify the longest cycle and arrange the nodes within it in a regular polygon to improve structural clarity. 
The nodes outside this cycle are then positioned using a layered layout, considering their connections to the nodes within the cycle. 
This initial layout ensures structural coherence across nesting levels, facilitating the clear representation of sentence-level semantic structures, while maintaining alignment with the top-to-bottom, left-to-right reading order of the text. 

\subsubsection{Force-Directed Layout Algorithm}
Contextual connections within the text can lead to the integration of nested graphs from different sentences, resulting in a more intricate structure, where some nodes may share common substructures, and links can span multiple nesting levels.
This complexity makes it challenging to maintain clarity and readability in the visualization. 
To address this challenges, we employ a force-directed layout algorithm that iteratively refines the initial layout, optimizing the spatial distribution of nodes to enhance structural clarity and align with the natural reading order. 

\begin{figure}[t]
  \centering
  \includegraphics[width=0.6\linewidth]{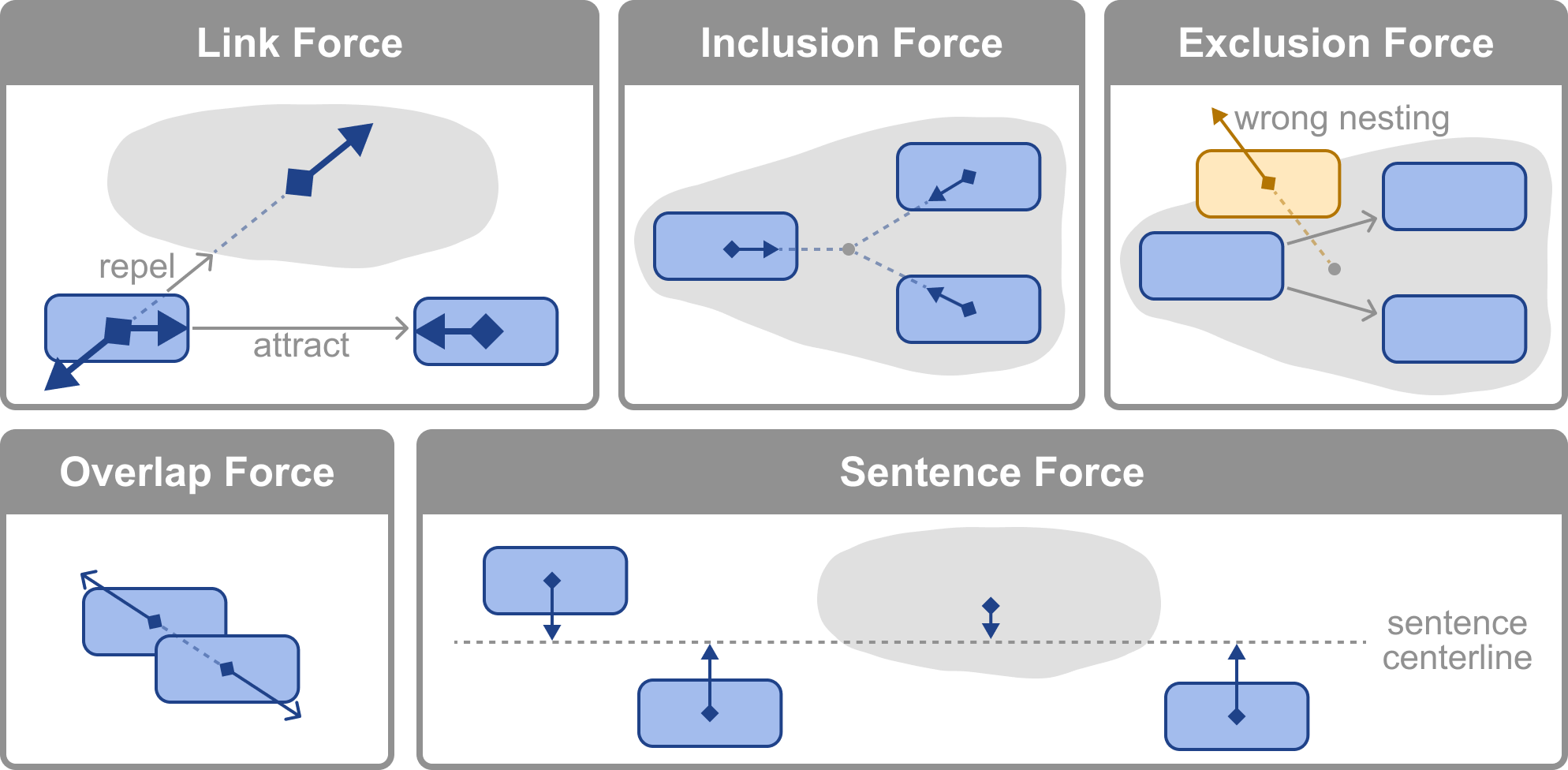}
  \caption{A schematic diagram of the five forces defined in the force-directed process}
  \label{fig:force}
\end{figure}

As shown in \autoref{fig:force},we define the following five types of forces in the force-directed layout algorithm used in \system{}.

\begin{itemize}[leftmargin=*]
\item \textbf{Link Force} acts as a spring between connected nodes, regardless of their nesting levels, maintaining an ideal distance by attracting or repelling them based on their deviation from a predefined optimal length.
\item \textbf{Inclusion Force} ensures that nodes nested within the same composite node remain spatially cohesive by exerting a constant force toward the center, reinforcing their semantic association.
\item \textbf{Exclusion Force} acts on nodes that don't structurally belong to a composite node but are spatially enclosed within its boundary. These nodes experience a constant repulsive force pushing them away from the composite node's center, preventing misinterpretation of nested structure.
\item \textbf{Overlap Force} prevents overlapping between atomic nodes by exerting a mutual repulsive force, ensuring clear visual separation and enhancing readability.
\item \textbf{Sentence Force} pulls each node toward the horizontal centerline of the sentences they belong to, reducing vertical dispersion within the same sentence while maintaining clear separation between different sentences. It helps to align with the top-to-bottom reading order, enhancing readability and structural clarity.
\end{itemize}

During each iteration of the force-directed layout process, all nodes, including both atomic and composite nodes, are subject to a net force, which is computed and used to adjust their positions accordingly. In particular, when a composite node moves, all nodes contained within it are displaced accordingly to maintain structural consistency. To improve visual clarity, the positions of atomic nodes are discretized, aligning them to fixed intervals. This iterative process continues until the positions of all nodes stabilize or a predefined maximum number of iterations is reached, at which point the final layout is determined.

\begin{figure*}[t]
  \centering
  \includegraphics[width=\textwidth]{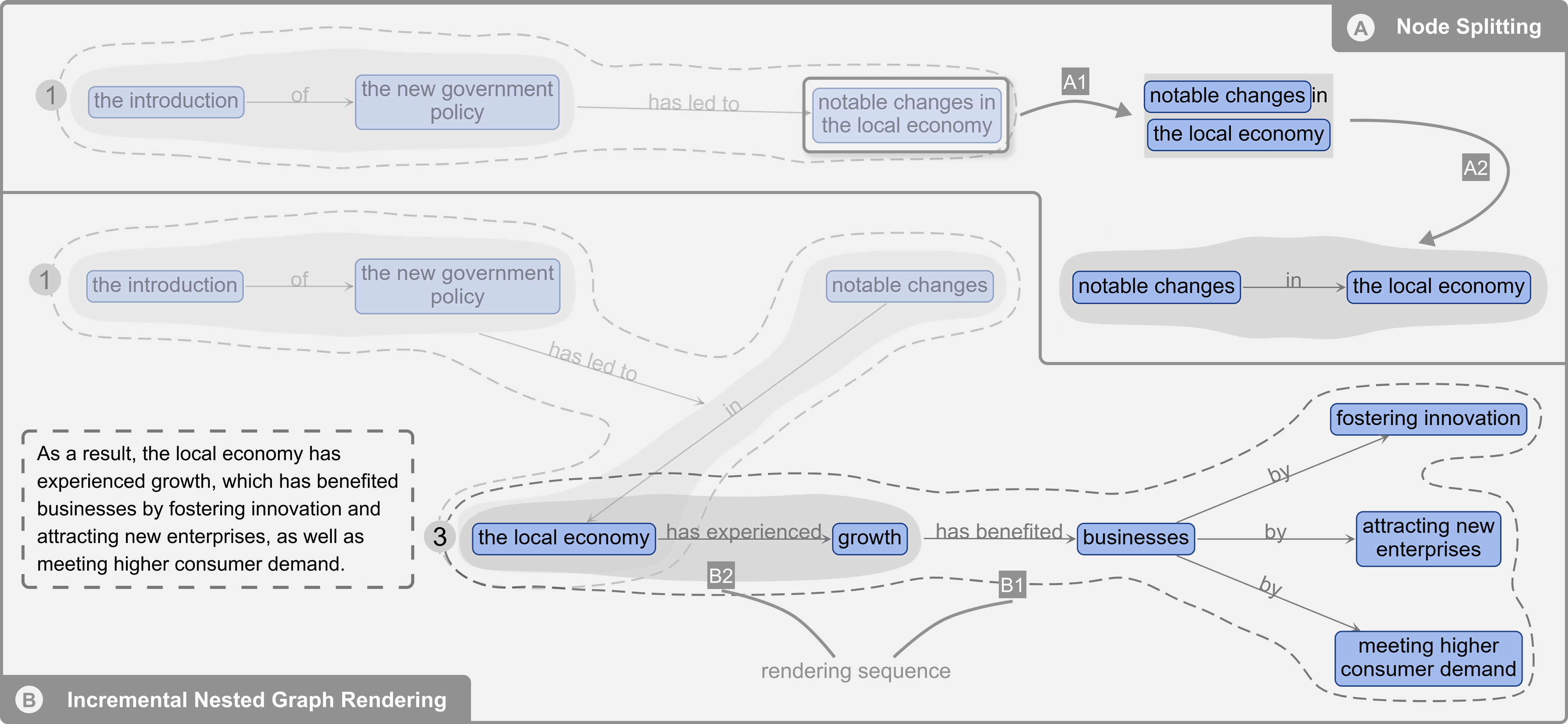}
  \caption{The Process of Progressive Presentation with Animation: Before a sentence mentioning ``the local economy'' appears, the relevant node, ``notable changes in the local economy'', is first decomposed (A) and moved to the region where the new sentence will appear. Subsequently, the new sentence's nodes and edges are progressively rendered from outside to inside (B).}
  \label{fig:progressive}
\end{figure*}

\subsection{Progressive Rendering Strategy}
Presenting all information at once risks overwhelming users and obscuring key insights. The interwoven nested relationships and densely connected nodes make it difficult to follow logical connections without a structured, step-by-step approach. By progressively revealing information, the presentation remains clear and manageable, allowing users to engage with the information incrementally while avoiding cognitive overload, as shown in \autoref{fig:progressive}.

\textbf{Click-Driven Sentence Presentation}. To accommodate different reading speeds, our system progressively presents information in sync with user interaction. By clicking on the interface, the next sentence and its corresponding visualization are introduced incrementally, ensuring a smooth and controlled information flow. To further enhance focus on newly introduced content, the visibility of existing elements in the canvas is reduced before the next sentence appears, minimizing visual clutter and preventing distractions. 

\textbf{Node Splitting}. In some cases, when a newly introduced sentence shares common sub-concepts with previously mentioned entities, it triggers the structural decomposition of prior nodes to establish new connections. This process visually reinforces contextual connections through node splitting. Specifically, the node to be split first changes color to match its resulting composite node, with its internal nodes encircled (\autoref{fig:progressive}.A1). The original bounding box then gradually reshapes into an enclosing structure, while the internal nodes shift into their new positions, forming the appropriate links (\autoref{fig:progressive}.A2).

\textbf{Layout Adjustment}. To align with the natural left-to-right reading order, newly introduced sentence is arranged according to the initial layout strategy described in Section~\ref{sec:layout}. If an entity in the new sentence overlaps with one from the previous content, its corresponding node is fixed in the new position to reinforce contextual continuity and the positions of other nodes are then optimized iteratively using a force-directed layout algorithm described in Section~\ref{sec:layout}.

\textbf{Incremental Nested Graph Rendering}. The nested graph of the new sentence is revealed incrementally, following the reading order of the original text. The rendering progresses from outer to inner structures and from left to right, first introducing composite nodes and then sequentially rendering their internal nodes and edges. This structured presentation ensures a clear and coherent flow of information, allowing users to first grasp high-level relationships before exploring finer details. 

\textbf{Column-Based Graph Arrangement}. Sentences without overlapping entities are organized into separate columns when presenting their corresponding nested graphs, preserving the multiple perspectives inherent in the original text while reducing visual clutter caused by increasing text length. Additionally, during progressive rendering, this structured arrangement helps users clearly recognize shifts in thematic focus or key information, enhancing their comprehension of the gradually unfolding content.
\section{\system{} Interactions}

\system{} provides interactions that support flexible exploration of sentences and entities. \textbf{Context-aware bidirectional linking} enables seamless navigation between the visualization and the original text \textbf{(D4)}, while \textbf{entity-based graph review} supports post-reading review and summarization around salient entities and their contexts \textbf{(D5)}.

\subsection{Context-Aware Bidirectional Linking}
\begin{figure*}[t]
  \centering
  \includegraphics[width=\textwidth]{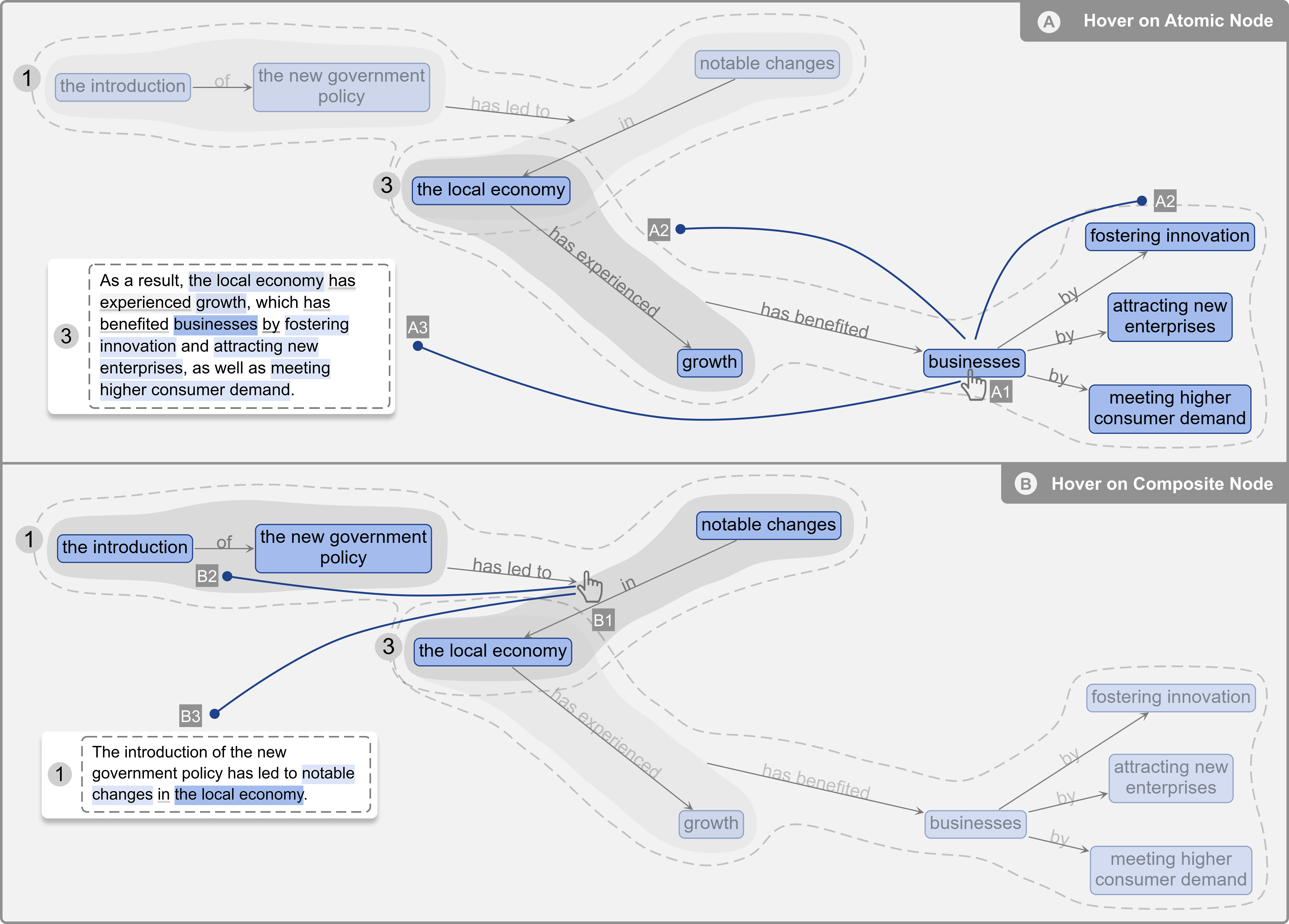}
  \caption{When the user hovers over a node, whether it's an atomic node (A1) or a composite node (B1), the node and its connected nodes (A2/B2) are prominently highlighted, and the corresponding text is also highlighted (A3/B3).}
  \label{fig:highlight}
\end{figure*}

Given the structural complexity of the graph, highlighting a single node alone may not provide enough context. To address this, our system reduces the visibility of unrelated nodes and links when a user hovers over any node, whether atomic or composite, ensuring that only relevant connections remain prominent. This selective emphasis helps users focus on meaningful relationships without being overwhelmed by extraneous details.

Additionally, when exploring nested graphs, users may need to refer back to the original text to understand implicit connections that may not be fully captured visually or to verify the entities and relationships. As shown in \autoref{fig:highlight}, hovering over a node in the nested graph will not only highlight its relevant nodes or links, but also its corresponding mentions in the text, allowing users to quickly locate relevant passages and confirm contextual details.

On the other hand, as users engage with detailed explanations in the original text, they may need to refer to the nested graphs for a structured representation of complex entity relationships. To support this, hovering over words in the text highlights not only their corresponding nodes in the graph but also the related nodes and links, providing a clearer view of their connections. This allows users to locate concepts of interest within the graph while clearly understanding their relationships in context.

\subsection{Entity-based Graph Review}

\begin{figure*}[t]
  \centering
  \includegraphics[width=\textwidth]{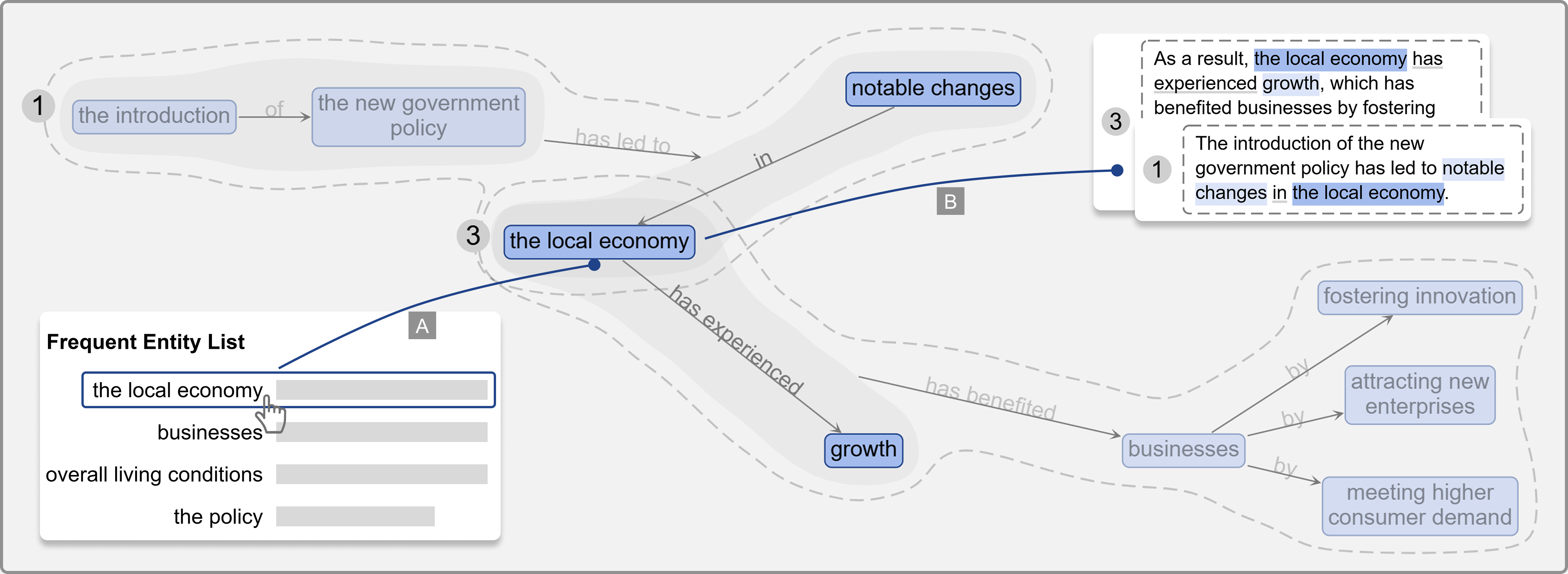}
  \caption{When a user clicks on an entry of the frequent entity list, the corresponding node and its connected nodes are highlighted in the graph (A), while the relevant content in the original text is also highlighted (B).}
  \label{fig:frequent}
\end{figure*}

To support effective information review, \system{} provides a ranked list of frequent entities extracted from the nested graph, enabling users to explore the graph based on key entities. Entities are sorted by the connectivity of their corresponding nodes in the graph. Specifically, the degree of each composite node is propagated to all of its contained atomic nodes, helping users to quickly identify salient entities that play a central role in the text. As shown in \autoref{fig:frequent}, clicking on a high-frequency entity from the list allows users to quickly locate it within the nested graph structure. And the interface highlights the selected node along with its directly connected nodes, clearly revealing its contextual relevance. This interaction design helps users efficiently explore key entities and their contexts, enhancing the overall review experience.
\section{User Evaluation}
To evaluate whether \system{} effectively supports medium-length knowledge-intensive text comprehension, we conducted a within-subject study with 12 participants.
In this study, we aim to answer the following research questions:
\begin{itemize}
    \item \textbf{RQ1}: Does \system{} improve the efficiency and quality of knowledge-intensive text comprehension?
    \item \textbf{RQ2}: Does the nested graph representation improve structural clarity compared to a single-layer graph?
    \item \textbf{RQ3}: Does the progressive rendering strategy with animation improve continuity in comprehension compared to static display?
\end{itemize}

Together, these research questions cover both the overall effectiveness and the specific design features of \system{}. RQ1 serves as an overall evaluation of comprehension support, while RQ2 and RQ3 focus on assessing the contributions of the two main features of \system{}, which are the nested graph representation and the progressive rendering strategy.

\subsection{Participants}
We recruited 12 participants (P1---P12, 8 males and 4 females), including undergraduate and graduate students. 
All participants were fluent English readers without prior background or interest in the study materials and a majority (9 out of 12) reported daily engagement in knowledge-intensive comprehension tasks, including scientific literature, academic textbooks, and in-depth news reports.

\subsection{Comparison Baselines}

To evaluate \system{}, we selected two baselines that allow us to evaluate the effectiveness of its two main components: (1) a novel \textbf{nested graph representation} for information abstraction, and (2) a \textbf{progressive rendering strategy with animation} that supports incremental comprehension.

\textbf{Graphologue}~\cite{Graphologue} serves as a baseline to evaluate the effectiveness of the nested graph representation. It is a representative prior work that transforms text into single-layer graphical diagrams to support information-seeking and question-answering tasks. By comparing with it, we assess whether introducing a nested graph representation can provide clearer structural support for knowledge-intensive text comprehension \textbf{(RQ2)}. 

\textbf{Static Display} is another baseline designed to evaluate the contribution of the progressive rendering strategy. It is a variant of our system that retains the same nested graph representation but presents it all at once without progressive rendering and animation. This comparison allows us to isolate the effect of progressive rendering on improving continuity in comprehension \textbf{(RQ3)}.

Here, we did not include a plain text baseline in our study, as prior work ~\cite{textBaseline1,textBaseline2} has already shown that structured visualizations substantially improve comprehension compared to plain text, and the original text was always accessible in all experimental conditions. Therefore, we focused on comparing different forms of visual representations to better understand the contributions of nested structure and progressive rendering.

\subsection{Tasks}

To evaluate users' comprehension of knowledge-intensive texts, we designed reading and question-answering tasks based on medium-length passages and open-ended questions adapted from standard reading comprehension datasets to balance complexity and feasibility. We used three passages from distinct domains of policy analysis, climate change, and the medicine industry. Each passage consisted of 6–8 sentences (approximately 130 words) and was characterized by dense knowledge content. Each passage was paired with three open-ended questions covering factual, causal, and summarization types. A majority of these questions (6 out of 9) required linking information across multiple sentences, thereby testing participants' ability to connect interrelated concepts within a broader context.

\subsection{Procedure}

Each study consisted of three sessions, one for each system, and the system order was counterbalanced across participants to mitigate learning effects. Each session began with an introduction to the assigned system using a sample passage excluded from the main task materials, which served to illustrate the visualization design and interactive functionalities. Participants were then given up to three minutes to read and comprehend a knowledge-intensive passage using the system, during which their reading time was recorded. This limit was determined through a pilot study and proved sufficient, as no participants exceeded it in the main study. After the reading period, participants answered three questions related to the passage, with the option to revisit both the text and the visualization. Their response time for each question was also recorded.

After completing all sessions, participants were asked to fill out a post-study questionnaire designed to assess their perceptions of each system's effectiveness in supporting knowledge-intensive text comprehension. For each system, participants rated their level of agreement with various statements on a 5-point Likert scale. 
Subsequently, we conducted an interview in which participants were encouraged to compare the three systems, reflect on their experiences, and freely discuss perceived strengths, weaknesses, and suggestions for improvement. This was intended to provide deeper insights into user preferences and to identify potential usability issues that might not have been captured by the questionnaire.

\subsection{User Performance}
To evaluate and compare the effectiveness of the three systems in supporting knowledge-intensive text comprehension, we examined three key performance metrics: (1) \textbf{reading time}, (2) \textbf{response time}, and (3) \textbf{response quality}. These metrics capture different stages of the comprehension process, from initial information processing to task-driven recall and integration. The within-subject design minimized the influence of individual differences in reading and comprehension performance.
After collecting the data, we conducted paired t-tests to compare \system{} with the two baseline systems. We first performed a Shapiro-Wilk test at a significance level of 0.05 to check the normality of the paired differences. Given that the differences followed a normal distribution, we calculated the mean and standard deviation for all sets, along with the t-value and p-value for the comparison. 

\begin{figure}[t]
  \centering
  \includegraphics[width=0.6\linewidth]{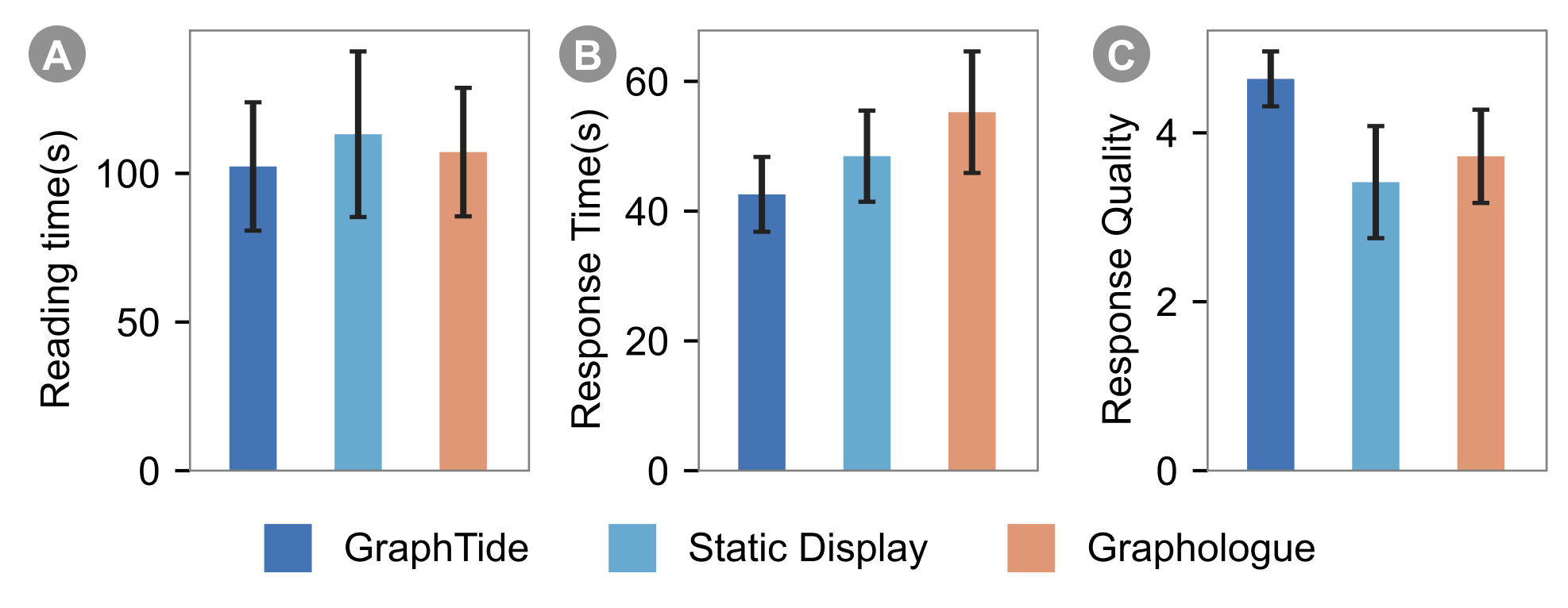}
  \centering
  \caption{User Performance measured using three quantitative metrics: (A), (B) and (C) shows the mean values of reading time, response time and response quality respectively, with error bars indicating the 95\% confidence interval.}
  \label{fig:performance}
\end{figure}


\textbf{Reading Time.} 
We measured participants' reading time under each system as an indicator of reading fluency and initial processing effort. As shown in \autoref{fig:performance}.A, participants spent on average 101.6s (SD = 40.3) reading with \system{}, compared to 113.0s (SD = 32.6) with Static Display and 107.2s (SD = 48.1) with Graphologue. 

\textbf{Response Time.}
We measured participants' response time per question as an indicator of task-driven information retrieval speed in the post-reading phase. As shown in \autoref{fig:performance}.B, participants spent on average 42.6s (SD = 16.8) with \system{} answering each question, compared to 48.5s (SD = 27.3) with Static Display and 55.3s (SD = 20.5) with Graphologue. 
A paired t-test showed that response time in the post-reading phase was significantly shorter with \system{} than with Graphologue (t = -2.34, p < 0.05).

\textbf{Response Quality.}
To assess participants' quality of comprehension, we developed a 5-point scoring standard for each question and evaluated each participant's response accordingly. This allowed us to quantify answer quality in terms of completeness, accuracy, and relevance. As shown in \autoref{fig:performance}.C, on average, participants scored 4.64 (SD = 0.95) with \system{}, compared to 3.42 (SD = 1.61) with Static Display and 3.72 (SD = 1.93) with Graphologue. 
Paired t-tests showed that the response quality with \system{} was significantly higher compared to Graphologue ( t = 2.90, p < 0.05) and Static Display (t = 3.36, p < 0.05).

Overall, although reading time was comparable across conditions, participants provided higher-quality responses in less time with \system{}. This suggests that \system{} enhances the efficiency and quality of knowledge-intensive text comprehension \textbf{(RQ1)}.

To better understand the performance differences, we analyzed user behavior and feedback to examine how the systems supported users in locating and linking pieces of relevant information during the question-answering phase.

In Graphologue, the single-layer representation, with many nodes containing long text segments, hindered users' ability to quickly locate the relevant information for a given question. Moreover, the diagram did not visually capture certain contextual relationships, making it difficult to trace connections across the text. 
In Static Display, although the nested structure was preserved, participants had not experienced its progressive construction and therefore lacked a clear mental model of how the content was built. This made it more difficult to navigate the graph confidently and efficiently to link relevant information. 

In contrast, the progressive construction of nested graphs in \system{} allowed users to follow the flow of information more naturally and understand how different parts were related. Based on a comprehensive mental model, users were able to more efficiently locate relevant information during the question-answering phase, with the aid of the interactive features of \system{}.

\begin{figure*}[t]
  \centering
  \includegraphics[width=\textwidth]{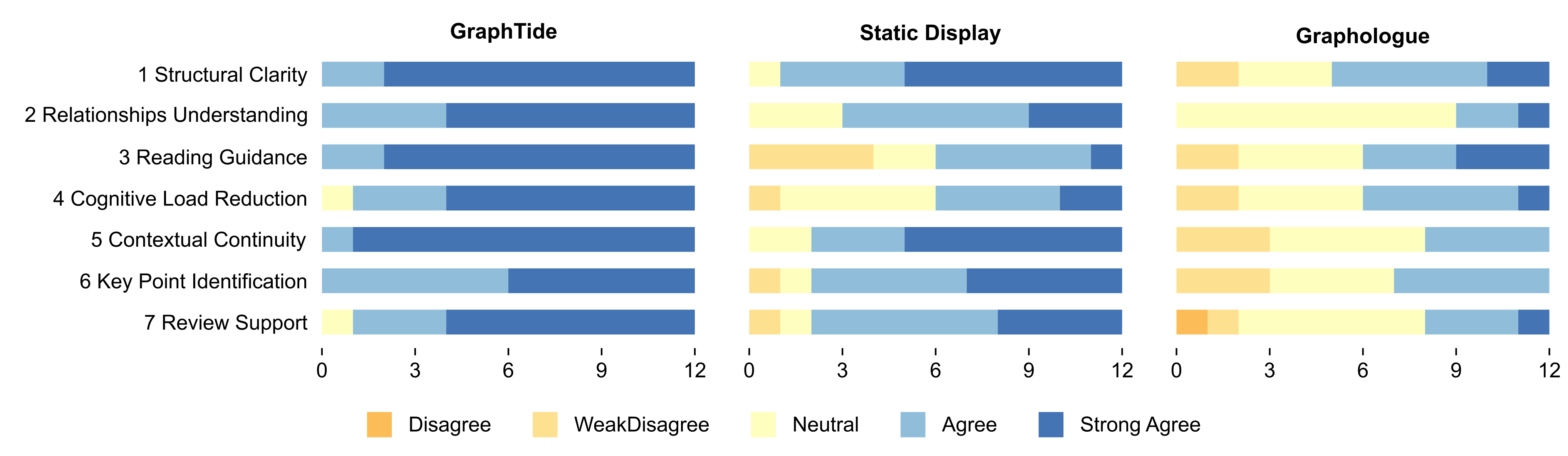}
  \caption{Participants' responses to various aspects of text comprehension support in \system{}, Static Display, and Graphologue, measured on a 5-point Likert scale.}
  \label{fig:feedback}
\end{figure*}

\subsection{User Feedback}
In the post-task questionnaire, participants were asked to rate their agreement with seven targeted statements on a 5-point Likert scale (1---strongly disagree, 5---strongly agree), in order to assess how well each system supported different aspects of text comprehension. The results were summarized in \autoref{fig:feedback}.

\subsubsection{\system{} Enhances Structural Clarity and Relationship Comprehension} 
Participants consistently reported that \system{} helped them better grasp the structure of information (\autoref{fig:feedback}.1) and enhanced their understanding of relationships between entities (\autoref{fig:feedback}.2).
Most participants (11/12) found that the nested graph in \system{} offered a clearer and more interpretable structure of the text \textbf{(RQ2)}. This was particularly helpful for sentences with complex structures or multiple interacting entities, where the visualization enabled users to quickly grasp how various elements were related. For instance, P10 noted that \textit{``without the nested structure, the diagram can't accurately express the meaning of complex sentences''}. Similarly, P4 commented that a flat graph \textit{``isn't able to capture the complexity of relationships.''} 
In contrast, the nested graph representation in \system{} allowed users to comprehend both structure and meaning directly from the graph. As several participants (P1, P7, P8, P10) observed, they were able to \textit{``capture the meaning entirely from the visualization''} without the need to refer back to the original text.

Additionally, the progressive rendering strategy was also considered helpful for understanding both sentence structure and inter-entity relationships. Most participants (10/12) pointed out that being presented with a dense and complex graph all at once made it more difficult to understand how entities were connected. In contrast, the progressive rendering preserved clarity throughout the reading process. As P3 described, \textit{``starting with less and adding more progressively makes it easier to understand how everything is connected.''} 

\subsubsection{\system{} Guides Reading and Reduces Cognitive Load} 
Participants founded that \system{} effectively guided the reading process (\autoref{fig:feedback}.3) and substantially reduced the cognitive effort (\autoref{fig:feedback}.4).

All participants except P9 noted that directly reading the original text was cognitively demanding due to domain-specific terminology and complex logical structures. Visualizing relationships of entities helped alleviate this burden. As P2 remarked, \textit{``the graph makes it much easier to see how different pieces of information are linked together.''}

However, participants also pointed out that visualizations can introduce cognitive overhead. In this regard, \system{} was seen as particularly effective, as its progressive rendering followed the natural order of the text and allowed users to read the diagram smoothly. Participants described the experience as \textit{``easy to follow''} (P8), \textit{``without additional burden''} (P2), and \textit{``just like following the passage itself''} (P7) . In contrast, Static Display forced users to locate the corresponding visualization for each sentence individually, which most participants (P1, P2, P5, P7, P8, P10, P11) described as a \textit{``tedious and distracting''} experience. With Graphologue, many struggled to reconstruct the full meaning from the flat graph and often had to return to the original text for clarification. As P4 noted, \textit{``I get lost in the middle of the diagram and have to go back to the paragraph to understand what it is talking about.''}

In addition, several participants (P1, P3, P7) noted that the animation in \system{} made the reading experience more engaging, whereas the static nested graph and the flat graph with layered layout felt visually dull. As P3 described, \textit{``the animation makes it feel less dry, and I actually want to keep reading to see how the next part connects.''}

\subsubsection{\system{} Enhances Contextual Continuity and Facilitates Information Integration} 

Many participants noted that the splitting animation and node transitions in \system{} effectively highlighted contextual continuity between sentences (\autoref{fig:feedback}.5). These visual cues helped users identify how the current content was connected to previously introduced entities. As P4 remarked, \textit{``when a node moves into the new sentence, I know that it is referring back to the previous content.''} Similarly, P7 explained, \textit{``splitting shows me how the same entity plays different roles across sentences.''} P9 also noted that \textit{``it is easy to track recurring concepts because the graph visually brings them together.''}

This emphasis on contextual continuity not only supported comprehension but also facilitated the linking of relevant information across sentences. In our study, the question-answering task that followed each reading phase simulated the cognitive process of information retrieval across sentences. \system{} allowed participants to retrieve and link information more effectively, as evidenced by the shortest average response time and the highest response quality. As P6 noted, \textit{``when answering questions, it is easy to find the answer. Once I locate the relevant node, I can clearly see all the connected ones.''}

Participants emphasized that this ability to visually trace and link related information across the sentences not only improved task performance, but also reduced the effort needed to piece together the ideas. Several users noted that they \textit{``don't need to jump back and forth between sentences''} (P2, P8), and instead could \textit{``see how everything fit together at a glance''} (P5).

\subsubsection{\system{} Facilitates Information Review and Summarization} 
Participants reported that \system{} was effective in identifying key information (\autoref{fig:feedback}.6) and supported the review and summarization process (\autoref{fig:feedback}.7) through a combination of visual and interactive design features.

Several participants (P1, P3, P6, P10) observed that key entities stood out on the graph because they were connected by multiple lines or enclosed within several groupings. This visual prominence \textit{``makes it easier to know what to focus on when looking back''} (P5).

Additionally, the frequent entity list also provided a useful entry point, particularly during the review phase. P11 remarked, \textit{``I can find a word that I'm interested in, or that I think represents the theme of the passage, and immediately jump to the related part of the graph to see what is connected to it.''}

Most participants (10/12) noted that the progressive rendering helped reinforce their memory of the graph's structure, which in turn made it easier to locate relevant information later. As P4 described, \textit{``seeing the graph build up step by step makes it easier to remember how things are connected''} and P5 added that the gradual exposure made the content \textit{``stick better in memory.''}

Finally, participants appreciated the multi-column layout, which often reflected different topical segments. As P7 described, \textit{``each column feels like a different topic. It helps me mentally organize what the passage is about.''}
\section{Discussion}
The user evaluation results highlight the effectiveness, limitations, and opportunities of \system{}. We discuss these aspects below.

\textbf{Incorporating discourse-level semantic structures.}
While \system{} effectively captures sentence-level structures and highlights shared entities across sentences, it currently lacks representations of discourse-level semantics, such as causality, contrast, elaboration, that are beneficial for deeper comprehension. 
Several participants noted that while the nested graph effectively clarified complex inter-entity relationships and their co-occurrences across sentences, it did not sufficiently capture the implicit logical relations that connect these concepts at the discourse level. 
For instance, while the graph shows that ``industrial activities contribute to emissions'' and ``policies aim to reduce emissions'' both refer to the concept of emissions, it does not convey the underlying problem-solution relationship implied between the two statements. 

To support more advanced reasoning and sensemaking, future work could incorporate discourse annotation frameworks such as Rhetorical Structure Theory (RST)~\cite{RST} or Segmented Discourse Representation Theory (SDRT)~\cite{SDRT} into the graph representation. This would allow visualizations to encode not only entity-level links, but also higher-level rhetorical relations between discourses. By making such discourse structures explicit, the system could further assist users in tracing argumentative flows, identifying supporting evidence, and interpreting logical dependencies with greater clarity.

\textbf{Expanding graph interactions and summarization support.}
While \system{} enables users to explore the graph presentation through features such as navigating back to the original text and locating context by frequent entities. 
In the user study, several participants expressed interest in more goal-oriented features to better support downstream tasks. 
In particular, they wished for the ability to bookmark important nodes for later review or comparison, especially when working with long or complex documents. 
Others suggested that generating summaries of selected subgraphs, such as clusters of nodes related to a specific topic, could help distill key ideas while maintaining their contextual grounding. 

Additionally, the system could be extended with more analytical features, such as annotating nodes with sentiment labels or classifying them by rhetorical roles. 
These capabilities could transform the graph from a comprehension aid into an interactive space for interpretation and reflection. 
Enhancing interactivity in this direction may further support personalized reading strategies and promote deeper, more structured sensemaking. 

\textbf{Supporting human-centered narrative progression.}
To enhance user engagement and comprehension, human-centered narrative progression can be integrated by customizing the reading flow to align with individual preferences and needs. 
For instance, users may wish to focus on specific entities or topics, skip over familiar sections, or replay key transitions to reinforce understanding. 
Offering such flexibility allows the system to tailor the reading experience to different user goals, whether for quick skimming or deeper analysis.

Human-centered progression can also be facilitated through adaptive features that respond to individual user behavior. 
For example, the system could highlight relevant entities based on the user’s current focus or suggest important passages they might have missed. 
This customization improves comprehension flow, making it easier for users to navigate complex, knowledge-intensive texts while concentrating on the most critical content.
By enabling users to control how information is presented and revisited, the system fosters a more engaging and efficient reading experience, adapting to both the complexity of the text and the user's evolving understanding.

\textbf{Investigating cognitive impact of animation styles.}
The use of animation in \system{} plays a crucial role in supporting comprehension by facilitating the progressive presentation of information. 
However, the precise cognitive effects of different animation styles on user engagement and information processing remain a subject for further exploration. 
Various animation styles, such as fading, morphing, or semantic zooming, can influence how users perceive and interact with the visualized content. 

Preliminary feedback from the user study suggests that both smooth, continuous transitions and splitting animations play key roles in enhancing user engagement and comprehension. 
However, some users may prefer certain animation styles over others based on personal preferences or task-specific needs. 
For instance, morphing animations could emphasize the transformation of key concepts, whereas fading might be more effective in reducing visual clutter, allowing users to focus on relevant information without distraction.
Further research could examine how different animation styles influence cognitive load, attention retention, and information integration. 
A deeper understanding of these effects could inform the design of future systems, enabling the use of tailored animations to enhance specific aspects of comprehension, such as supporting memory retention or facilitating deeper analytical reasoning.
\section{Conclusion}
In this paper, we introduced \system{}, a visualization technique for text augmentation that progressively constructs nested entity-relationship graphs with animation to support comprehension of knowledge-intensive text. 
Our method leverages an on-demand decomposition pipeline to construct nested graph representation and a structure-aware layout optimization algorithm to ensure the spatial synchronization of the graph and text. 
By incrementally revealing sentences and entities through smooth and continuous transitions, \system{} helps users maintain context and construct mental models gradually throughout the reading process. 
A user study demonstrates that \system{} significantly improves comprehension compared to traditional graph-based techniques and static nested graph representations.

\bibliographystyle{ACM-Reference-Format}
\bibliography{citation}

@article{GPT-4o,
    title={GPT-4o System Card}, 
    author={OpenAI},
    year    = {2024},
    journal = {arXiv preprint arXiv: 2410.21276}
}

@inproceedings{Graphologue,
    author = {Jiang, Peiling and Rayan, Jude and Dow, Steven P. and Xia, Haijun},
    title = {Graphologue: Exploring Large Language Model Responses with Interactive Diagrams},
    year = {2023},
    doi = {10.1145/3586183.3606737},
    booktitle = {Proceedings of the 36th Annual ACM Symposium on User Interface Software and Technology},
    articleno = {3},
    numpages = {20},
}

@ARTICLE{KNowNEt,
  author={Yan, Youfu and Hou, Yu and Xiao, Yongkang and Zhang, Rui and Wang, Qianwen},
  journal={IEEE Transactions on Visualization and Computer Graphics}, 
  title={KNowNEt:Guided Health Information Seeking from LLMs via Knowledge Graph Integration}, 
  year={2025},
  volume={31},
  number={1},
  pages={547-557},
  doi={10.1109/TVCG.2024.3456364}
}

@ARTICLE{KG4Vis,
  author={Li, Haotian and Wang, Yong and Zhang, Songheng and Song, Yangqiu and Qu, Huamin},
  journal={IEEE Transactions on Visualization and Computer Graphics}, 
  title={KG4Vis: A Knowledge Graph-Based Approach for Visualization Recommendation}, 
  year={2022},
  volume={28},
  number={1},
  pages={195-205},
  doi={10.1109/TVCG.2021.3114863}
}

@ARTICLE{KGScope,
  author={Hsuan Yuan, Chao-Wen and Yu, Tzu-Wei and Pan, Jia-Yu and Lin, Wen-Chieh},
  journal={IEEE Transactions on Visualization and Computer Graphics}, 
  title={KGScope: Interactive Visual Exploration of Knowledge Graphs With Embedding-Based Guidance}, 
  year={2024},
  volume={30},
  number={12},
  pages={7702-7716},
  doi={10.1109/TVCG.2024.3360690}
}

@ARTICLE{CommonsenseVIS,
  author={Wang, Xingbo and Huang, Renfei and Jin, Zhihua and Fang, Tianqing and Qu, Huamin},
  journal={IEEE Transactions on Visualization and Computer Graphics}, 
  title={CommonsenseVIS: Visualizing and Understanding Commonsense Reasoning Capabilities of Natural Language Models}, 
  year={2024},
  volume={30},
  number={1},
  pages={273-283},
  doi={10.1109/TVCG.2023.3327153}
}

@article{EffectOfText,
    author = {Kuo-En Chang, Yao-Ting Sung and Ine-Dai Chen},
    title = {The Effect of Concept Mapping to Enhance Text Comprehension and Summarization},
    journal = {The Journal of Experimental Education},
    volume = {71},
    number = {1},
    pages = {5--23},
    year = {2002},
    doi = {10.1080/00220970209602054},
}

@article{ReviewOfTextVisualization,
author = {Nualart Vilaplana, Jaume and Pérez-Montoro, Mario and Whitelaw, Mitchell},
year = {2014},
month = {05},
pages = {221-235},
title = {How we draw texts: A review of approaches to text visualization and exploration},
volume = {23},
journal = {El Profesional de la Informacion},
doi = {10.3145/epi.2014.may.02}
}

@article{MemoryCapacity,
author = {Schurer, Teresa and Opitz, Bertram and Schubert, Torsten},
year = {2020},
month = {03},
pages = {},
title = {Working Memory Capacity but Not Prior Knowledge Impact on Readers' Attention and Text Comprehension},
volume = {5},
journal = {Frontiers in Education},
doi = {10.3389/feduc.2020.00026}
}

@inproceedings{MultiDocument_Graph,
  author       = {Wei Li and
                  Xinyan Xiao and
                  Jiachen Liu and
                  Hua Wu and
                  Haifeng Wang and
                  Junping Du},
  title        = {Leveraging Graph to Improve Abstractive Multi-Document Summarization},
  booktitle    = {Proceedings of the 58th Annual Meeting of the Association for Computational
                  Linguistics, {ACL} 2020, Online, July 5-10, 2020},
  pages        = {6232--6243},
  publisher    = {Association for Computational Linguistics},
  year         = {2020},
  doi          = {10.18653/V1/2020.ACL-MAIN.555},
}

@article{SoS_TextVis,
  author       = {Mohammad Alharbi and
                  Robert S. Laramee},
  title        = {SoS TextVis: An Extended Survey of Surveys on Text Visualization},
  journal      = {Comput.},
  volume       = {8},
  number       = {1},
  pages        = {17},
  year         = {2019},
  doi          = {10.3390/COMPUTERS8010017},
}

@article{TextHighlight1,
  author       = {Matthew Brehmer and
                  Stephen Ingram and
                  Jonathan Stray and
                  Tamara Munzner},
  title        = {Overview: The Design, Adoption, and Analysis of a Visual Document
                  Mining Tool for Investigative Journalists},
  journal      = {{IEEE} Trans. Vis. Comput. Graph.},
  volume       = {20},
  number       = {12},
  pages        = {2271--2280},
  year         = {2014},
  doi          = {10.1109/TVCG.2014.2346431},
}

@article{TextHighlight2,
  author       = {Matthew Berger and
                  Katherine McDonough and
                  Lee M. Seversky},
  title        = {cite2vec: Citation-Driven Document Exploration via Word Embeddings},
  journal      = {{IEEE} Trans. Vis. Comput. Graph.},
  volume       = {23},
  number       = {1},
  pages        = {691--700},
  year         = {2017},
  doi          = {10.1109/TVCG.2016.2598667},
}

@inproceedings{TextHighlight3,
  author       = {Shimei Pan and
                  Michelle X. Zhou and
                  Yangqiu Song and
                  Weihong Qian and
                  Fei Wang and
                  Shixia Liu},
  title        = {Optimizing temporal topic segmentation for intelligent text visualization},
  booktitle    = {18th International Conference on Intelligent User Interfaces, {IUI}
                  2013, Santa Monica, CA, USA, March 19-22, 2013},
  pages        = {339--350},
  publisher    = {{ACM}},
  year         = {2013},
  url          = {https://doi.org/10.1145/2449396.2449441},
  doi          = {10.1145/2449396.2449441},
}

@article{WordCloud1,
  author       = {Weiwei Cui and
                  Yingcai Wu and
                  Shixia Liu and
                  Furu Wei and
                  Michelle X. Zhou and
                  Huamin Qu},
  title        = {Context-Preserving, Dynamic Word Cloud Visualization},
  journal      = {{IEEE} Computer Graphics and Applications},
  volume       = {30},
  number       = {6},
  pages        = {42--53},
  year         = {2010},
  url          = {https://doi.org/10.1109/MCG.2010.102},
  doi          = {10.1109/MCG.2010.102},
}

@inproceedings{WordCloud2,
  author       = {Quim Castell{\`{a}} and
                  Charles Sutton},
  title        = {Word storms: multiples of word clouds for visual comparison of documents},
  booktitle    = {23rd International World Wide Web Conference, {WWW} '14, Seoul, Republic
                  of Korea, April 7-11, 2014},
  pages        = {665--676},
  publisher    = {{ACM}},
  year         = {2014},
  doi          = {10.1145/2566486.2567977},
}

@article{WordCloud3,
  author       = {Ming{-}Te Chi and
                  Shih{-}Syun Lin and
                  Shiang{-}Yi Chen and
                  Chao{-}Hung Lin and
                  Tong{-}Yee Lee},
  title        = {Morphable Word Clouds for Time-Varying Text Data Visualization},
  journal      = {{IEEE} Trans. Vis. Comput. Graph.},
  volume       = {21},
  number       = {12},
  pages        = {1415--1426},
  year         = {2015},
  doi          = {10.1109/TVCG.2015.2440241},
}

@inproceedings{WordCloud4,
  author       = {Daniela Oelke and
                  Iryna Gurevych},
  title        = {A study on human-generated tag structures to inform tag cloud layout},
  booktitle    = {International Working Conference on Advanced Visual Interfaces, {AVI}
                  2014, Como, Italy, May 27-29, 2014},
  pages        = {297--304},
  publisher    = {{ACM}},
  year         = {2014},
  doi          = {10.1145/2598153.2598155},
}

@inproceedings{Chart1,
  author       = {Magdalena Jankowska and
                  Vlado Keselj and
                  Evangelos E. Milios},
  title        = {Relative N-gram signatures: Document visualization at the level of
                  character N-grams},
  booktitle    = {7th {IEEE} Conference on Visual Analytics Science and Technology,
                  {IEEE} {VAST} 2012, Seattle, WA, USA, October 14-19, 2012},
  pages        = {103--112},
  publisher    = {{IEEE} Computer Society},
  year         = {2012},
  doi          = {10.1109/VAST.2012.6400484},
}

@article{Chart2,
  author       = {Fernando Vieira Paulovich and
                  Rosane Minghim},
  title        = {HiPP: {A} Novel Hierarchical Point Placement Strategy and its Application
                  to the Exploration of Document Collections},
  journal      = {{IEEE} Trans. Vis. Comput. Graph.},
  volume       = {14},
  number       = {6},
  pages        = {1229--1236},
  year         = {2008},
  doi          = {10.1109/TVCG.2008.138},
}

@article{Chart3,
  author       = {Wei Chen and
                  Tianyi Lao and
                  Jing Xia and
                  Xinxin Huang and
                  Biao Zhu and
                  Wanqi Hu and
                  Huihua Guan},
  title        = {GameFlow: Narrative Visualization of {NBA} Basketball Games},
  journal      = {{IEEE} Trans. Multim.},
  volume       = {18},
  number       = {11},
  pages        = {2247--2256},
  year         = {2016},
  doi          = {10.1109/TMM.2016.2614221},
}

@article{Chart4,
  author       = {Danielle Albers Szafir and
                  D. Stuffer and
                  Y. Sohail and
                  Michael Gleicher},
  title        = {TextDNA: Visualizing Word Usage with Configurable Colorfields},
  journal      = {Comput. Graph. Forum},
  volume       = {35},
  number       = {3},
  pages        = {421--430},
  year         = {2016},
  doi          = {10.1111/CGF.12918},
}

@article{Graph1,
  author       = {Krist Wongsuphasawat and
                  David Gotz},
  title        = {Exploring Flow, Factors, and Outcomes of Temporal Event Sequences
                  with the Outflow Visualization},
  journal      = {{IEEE} Trans. Vis. Comput. Graph.},
  volume       = {18},
  number       = {12},
  pages        = {2659--2668},
  year         = {2012},
  doi          = {10.1109/TVCG.2012.225},
}

@article{Graph2,
  author       = {Yingcai Wu and
                  Shixia Liu and
                  Kai Yan and
                  Mengchen Liu and
                  Fangzhao Wu},
  title        = {OpinionFlow: Visual Analysis of Opinion Diffusion on Social Media},
  journal      = {{IEEE} Trans. Vis. Comput. Graph.},
  volume       = {20},
  number       = {12},
  pages        = {1763--1772},
  year         = {2014},
  doi          = {10.1109/TVCG.2014.2346920},
}

@article{Graph3,
  author       = {Florian Stoffel and
                  Wolfgang Jentner and
                  Michael Behrisch and
                  Johannes Fuchs and
                  Daniel A. Keim},
  title        = {Interactive Ambiguity Resolution of Named Entities in Fictional Literature},
  journal      = {Comput. Graph. Forum},
  volume       = {36},
  number       = {3},
  pages        = {189--200},
  year         = {2017},
  doi          = {10.1111/CGF.13179},
}

@article{Timeline1,
  author       = {Guodao Sun and
                  Yingcai Wu and
                  Shixia Liu and
                  Tai{-}Quan Peng and
                  Jonathan J. H. Zhu and
                  Ronghua Liang},
  title        = {EvoRiver: Visual Analysis of Topic Coopetition on Social Media},
  journal      = {{IEEE} Trans. Vis. Comput. Graph.},
  volume       = {20},
  number       = {12},
  pages        = {1753--1762},
  year         = {2014},
  doi          = {10.1109/TVCG.2014.2346919},
}

@article{Timeline2,
  author       = {Johanna Fulda and
                  Matthew Brehmer and
                  Tamara Munzner},
  title        = {TimeLineCurator: Interactive Authoring of Visual Timelines from Unstructured Text},
  journal      = {{IEEE} Trans. Vis. Comput. Graph.},
  volume       = {22},
  number       = {1},
  pages        = {300--309},
  year         = {2016},
  doi          = {10.1109/TVCG.2015.2467531},
}

@misc{Dagre,
  author = {Dagre},
  title = {Dagre: A Directed Graph Layout Library},
  year = {2015},
  howpublished = {\url{https://github.com/dagrejs/dagre}},
  note = {Accessed: 2025-02-21}
}

@article{Sugiyama,
  title={Methods for visual understanding of hierarchical system structures},
  author={Sugiyama, Kozo and Tagawa, Shojiro and Toda, Mitsuhiko},
  journal={IEEE Transactions on Systems, Man, and Cybernetics},
  volume={11},
  number={2},
  pages={109--125},
  year={1981},
  publisher={IEEE}
}

@article{ForceLayout1,
  title={Graph drawing by force-directed placement},
  author={Fruchterman, Thomas MJ and Reingold, Edward M},
  journal={Software: Practice and experience},
  volume={21},
  number={11},
  pages={1129--1164},
  year={1991},
}

@article{ForceLayout2,
  author       = {Tomihisa Kamada and
                  Satoru Kawai},
  title        = {An Algorithm for Drawing General Undirected Graphs},
  journal      = {Inf. Process. Lett.},
  volume       = {31},
  number       = {1},
  pages        = {7--15},
  year         = {1989},
  doi          = {10.1016/0020-0190(89)90102-6},
}

@article{ForceLayout3,
  author       = {Ashley Suh and
                  Mustafa Hajij and
                  Bei Wang and
                  Carlos Scheidegger and
                  Paul Rosen},
  title        = {Persistent Homology Guided Force-Directed Graph Layouts},
  journal      = {{IEEE} Trans. Vis. Comput. Graph.},
  volume       = {26},
  number       = {1},
  pages        = {697--707},
  year         = {2020},
  doi          = {10.1109/TVCG.2019.2934802},
}

@inproceedings{CompoundGraph_Definition,
  author       = {Kozo Sugiyama and
                  Kazuo Misue},
  editor       = {Franz{-}Josef Brandenburg},
  title        = {A Generic Compound Graph Visualizer/Manipulator: {D-ABDUCTOR}},
  booktitle    = {Graph Drawing, Symposium on Graph Drawing, {GD} '95, Passau, Germany,
                  September 20-22, 1995, Proceedings},
  series       = {Lecture Notes in Computer Science},
  volume       = {1027},
  pages        = {500--503},
  year         = {1995},
  doi          = {10.1007/BFB0021834},
}

@article{CompoundGraph_Layer_Layout1,
  author       = {Kozo Sugiyama and
                  Kazuo Misue},
  title        = {Visualization of structural information: automatic drawing of compound
                  digraphs},
  journal      = {{IEEE} Trans. Syst. Man Cybern.},
  volume       = {21},
  number       = {4},
  pages        = {876--892},
  year         = {1991},
  doi          = {10.1109/21.108304},
}

@inproceedings{CompoundGraph_Layer_Layout2,
  author       = {Kozo Sugiyama and
                  Kazuo Misue},
  editor       = {Franz{-}Josef Brandenburg},
  title        = {A Generic Compound Graph Visualizer/Manipulator: {D-ABDUCTOR}},
  booktitle    = {Graph Drawing, Symposium on Graph Drawing, {GD} '95, Passau, Germany,
                  September 20-22, 1995, Proceedings},
  series       = {Lecture Notes in Computer Science},
  volume       = {1027},
  pages        = {500--503},
  year         = {1995},
  doi          = {10.1007/BFB0021834},
}

@inproceedings{CompoundGraph_Layer_Layout3,
  author       = {Michael Forster},
  editor       = {Stephen G. Kobourov and
                  Michael T. Goodrich},
  title        = {Applying Crossing Reduction Strategies to Layered Compound Graphs},
  booktitle    = {Graph Drawing, 10th International Symposium, {GD} 2002, Irvine, CA,
                  USA, August 26-28, 2002, Revised Papers},
  series       = {Lecture Notes in Computer Science},
  volume       = {2528},
  pages        = {276--284},
  year         = {2002},
  doi          = {10.1007/3-540-36151-0\_26},
}

@inproceedings{CompoundGraph_Force_Layout1,
  author       = {Ugur Dogrus{\"{o}}z and
                  Erhan Giral and
                  Ahmet Cetintas and
                  Ali Civril and
                  Emek Demir},
  title        = {A Compound Graph Layout Algorithm for Biological Pathways},
  booktitle    = {Graph Drawing, 12th International Symposium, {GD} 2004, New York,
                  NY, USA, September 29 - October 2, 2004, Revised Selected Papers},
  series       = {Lecture Notes in Computer Science},
  volume       = {3383},
  pages        = {442--447},
  year         = {2004},
  doi          = {10.1007/978-3-540-31843-9\_45},
}

@article{CompoundGraph_Force_Layout2,
  author       = {Ugur Dogrus{\"{o}}z and
                  Erhan Giral and
                  Ahmet Cetintas and
                  Ali Civril and
                  Emek Demir},
  title        = {A layout algorithm for undirected compound graphs},
  journal      = {Inf. Sci.},
  volume       = {179},
  number       = {7},
  pages        = {980--994},
  year         = {2009},
  doi          = {10.1016/J.INS.2008.11.017},
}

@inproceedings{Sensecape,
  author       = {Sangho Suh and
                  Bryan Min and
                  Srishti Palani and
                  Haijun Xia},
  title        = {Sensecape: Enabling Multilevel Exploration and Sensemaking with Large
                  Language Models},
  booktitle    = {Proceedings of the 36th Annual {ACM} Symposium on User Interface Software
                  and Technology, {UIST} 2023, San Francisco, CA, USA, 29 October 2023-
                  1 November 2023},
  pages        = {1:1--1:18},
  year         = {2023},
  doi          = {10.1145/3586183.3606756}
}

@article{RST,
    title = {Rhetorical Structure Theory: Toward a functional theory of text organization},
    author = {WILLIAM C. MANN and SANDRA A. THOMPSON},
    pages = {243--281},
    volume = {8},
    number = {3},
    journal = {Text - Interdisciplinary Journal for the Study of Discourse},
    doi = {doi:10.1515/text.1.1988.8.3.243},
    year = {1988},
    lastchecked = {2025-04-08}
}

@book{SDRT,
  title={Logics of conversation},
  author={Asher, Nicholas and Lascarides, Alex},
  year={2003},
  publisher={Cambridge University Press}
}

@article{DeFT,
    author = {Ainsworth, Shaaron},
    year = {2006},
    month = {06},
    pages = {183-198},
    title = {DeFT: A conceptual framework for learning with multiple representations. Learning and Instruction, 16, 183-198},
    volume = {16},
    journal = {Learning and Instruction},
    doi = {10.1016/j.learninstruc.2006.03.001}
}

@misc{MindManager,
    author = {Mindjet},
    title = {MindManager – Intuitive Visualization Tools},
    howpublished = {Website},
    year = {1999},
    note = {\url{https://www.mindmanager.com/}},
    urldate = {2025-07-02},
}

@inproceedings{Sandbox,
  title={The Sandbox for analysis: concepts and methods},
  author={Wright, William and Schroh, David and Proulx, Pascale and Skaburskis, Alex and Cort, Brian},
  booktitle={Proceedings of the SIGCHI conference on Human Factors in computing systems},
  pages={801--810},
  year={2006}
}

@inproceedings{texSketch,
  author       = {Hariharan Subramonyam and
                  Colleen M. Seifert and
                  Priti Shah and
                  Eytan Adar},
  editor       = {Regina Bernhaupt and
                  Florian 'Floyd' Mueller and
                  David Verweij and
                  Josh Andres and
                  Joanna McGrenere and
                  Andy Cockburn and
                  Ignacio Avellino and
                  Alix Goguey and
                  Pernille Bj{\o}n and
                  Shengdong Zhao and
                  Briane Paul Samson and
                  Rafal Kocielnik},
  title        = {texSketch: Active Diagramming through Pen-and-Ink Annotations},
  booktitle    = {{CHI} '20: {CHI} Conference on Human Factors in Computing Systems,
                  Honolulu, HI, USA, April 25-30, 2020},
  pages        = {1--13},
  publisher    = {{ACM}},
  year         = {2020},
  url          = {https://doi.org/10.1145/3313831.3376155},
  doi          = {10.1145/3313831.3376155},
  bibsource    = {dblp computer science bibliography, https://dblp.org}
}

@inproceedings{LiquidText,
  author       = {Craig S. Tashman and
                  W. Keith Edwards},
  editor       = {Desney S. Tan and
                  Saleema Amershi and
                  Bo Begole and
                  Wendy A. Kellogg and
                  Manas Tungare},
  title        = {LiquidText: a flexible, multitouch environment to support active reading},
  booktitle    = {Proceedings of the International Conference on Human Factors in Computing
                  Systems, {CHI} 2011, Vancouver, BC, Canada, May 7-12, 2011},
  pages        = {3285--3294},
  publisher    = {{ACM}},
  year         = {2011},
  url          = {https://doi.org/10.1145/1978942.1979430},
  doi          = {10.1145/1978942.1979430},
  bibsource    = {dblp computer science bibliography, https://dblp.org}
}

@inproceedings{Jigsaw,
  author       = {John T. Stasko and
                  Carsten G{\"{o}}rg and
                  Zhicheng Liu and
                  Kanupriya Singhal},
  title        = {Jigsaw: Supporting Investigative Analysis through Interactive Visualization},
  booktitle    = {2nd {IEEE} Symposium on Visual Analytics Science and Technology, {IEEE}
                  {VAST} 2007, Sacramento, CA,USA, October 30 - November 1, 2007},
  pages        = {131--138},
  publisher    = {{IEEE} Computer Society},
  year         = {2007},
  url          = {https://doi.org/10.1109/VAST.2007.4389006},
  doi          = {10.1109/VAST.2007.4389006},
  bibsource    = {dblp computer science bibliography, https://dblp.org}
}

@inproceedings{EntityWorkspace,
  title={Entity workspace: an evidence file that aids memory, inference, and reading},
  author={Bier, Eric A and Ishak, Edward W and Chi, Ed},
  booktitle={Intelligence and Security Informatics: IEEE International Conference on Intelligence and Security Informatics, ISI 2006, San Diego, CA, USA, May 23-24, 2006. Proceedings 4},
  pages={466--472},
  year={2006},
  organization={Springer}
}

@inproceedings{Storytelling,
  author       = {Mahmud Shahriar Hossain and
                  Patrick Butler and
                  Arnold P. Boedihardjo and
                  Naren Ramakrishnan},
  editor       = {Qiang Yang and
                  Deepak Agarwal and
                  Jian Pei},
  title        = {Storytelling in entity networks to support intelligence analysts},
  booktitle    = {The 18th {ACM} {SIGKDD} International Conference on Knowledge Discovery
                  and Data Mining, {KDD} '12, Beijing, China, August 12-16, 2012},
  pages        = {1375--1383},
  publisher    = {{ACM}},
  year         = {2012},
  url          = {https://doi.org/10.1145/2339530.2339742},
  doi          = {10.1145/2339530.2339742},
  bibsource    = {dblp computer science bibliography, https://dblp.org}
}

@inproceedings{NetLens,
  author       = {Hyunmo Kang and
                  Catherine Plaisant and
                  Bongshin Lee and
                  Benjamin B. Bederson},
  editor       = {Pak Chung Wong and
                  Daniel A. Keim},
  title        = {NetLens: Iterative Exploration of Content-Actor Network Data},
  booktitle    = {1st {IEEE} Symposium On Visual Analytics Science And Technology, {IEEE}
                  {VAST} 2006, Baltimore, MD, USA, October 31 - November 2, 2006},
  pages        = {91--98},
  publisher    = {{IEEE} Computer Society},
  year         = {2006},
  url          = {https://doi.org/10.1109/VAST.2006.261426},
  doi          = {10.1109/VAST.2006.261426},
  bibsource    = {dblp computer science bibliography, https://dblp.org}
}

@misc{SentinelVisualizer,
  author = {{FMS Advanced Systems Group}},
  title = {Sentinel Visualizer},
  year = {2007},
  howpublished = {Website},
  note = {\url{https://sentinelvisualizer.com/}},
  urldate = {2025-07-07},
}

@inproceedings{GP-TSM,
  author       = {Ziwei Gu and
                  Ian Arawjo and
                  Kenneth Li and
                  Jonathan K. Kummerfeld and
                  Elena L. Glassman},
  editor       = {Florian 'Floyd' Mueller and
                  Penny Kyburz and
                  Julie R. Williamson and
                  Corina Sas and
                  Max L. Wilson and
                  Phoebe O. Toups Dugas and
                  Irina Shklovski},
  title        = {An AI-Resilient Text Rendering Technique for Reading and Skimming
                  Documents},
  booktitle    = {Proceedings of the {CHI} Conference on Human Factors in Computing
                  Systems, {CHI} 2024, Honolulu, HI, USA, May 11-16, 2024},
  pages        = {898:1--898:22},
  publisher    = {{ACM}},
  year         = {2024},
  url          = {https://doi.org/10.1145/3613904.3642699},
  doi          = {10.1145/3613904.3642699},
  bibsource    = {dblp computer science bibliography, https://dblp.org}
}

@article{PhraseNet,
  author       = {Frank van Ham and
                  Martin Wattenberg and
                  Fernanda B. Vi{\'{e}}gas},
  title        = {Mapping Text with Phrase Nets},
  journal      = {{IEEE} Trans. Vis. Comput. Graph.},
  volume       = {15},
  number       = {6},
  pages        = {1169--1176},
  year         = {2009},
  url          = {https://doi.org/10.1109/TVCG.2009.165},
  doi          = {10.1109/TVCG.2009.165},
  bibsource    = {dblp computer science bibliography, https://dblp.org}
}

@article{WordTree,
  author       = {Martin Wattenberg and
                  Fernanda B. Vi{\'{e}}gas},
  title        = {The Word Tree, an Interactive Visual Concordance},
  journal      = {{IEEE} Trans. Vis. Comput. Graph.},
  volume       = {14},
  number       = {6},
  pages        = {1221--1228},
  year         = {2008},
  url          = {https://doi.org/10.1109/TVCG.2008.172},
  doi          = {10.1109/TVCG.2008.172},
  bibsource    = {dblp computer science bibliography, https://dblp.org}
}

@article{SentenTree,
  author       = {Mengdie Hu and
                  Krist Wongsuphasawat and
                  John T. Stasko},
  title        = {Visualizing Social Media Content with SentenTree},
  journal      = {{IEEE} Trans. Vis. Comput. Graph.},
  volume       = {23},
  number       = {1},
  pages        = {621--630},
  year         = {2017},
  url          = {https://doi.org/10.1109/TVCG.2016.2598590},
  doi          = {10.1109/TVCG.2016.2598590},
  bibsource    = {dblp computer science bibliography, https://dblp.org}
}

@article{kintsch1978toward,
  title={Toward a model of text comprehension and production.},
  author={Kintsch, Walter and Van Dijk, Teun A},
  journal={Psychological review},
  volume={85},
  number={5},
  pages={363},
  year={1978},
  publisher={American Psychological Association}
}

@article{britton1991using,
  title={Using Kintsch's computational model to improve instructional text: Effects of repairing inference calls on recall and cognitive structures.},
  author={Britton, Bruce K and G{\"u}lg{\"o}z, Sami},
  journal={Journal of educational Psychology},
  volume={83},
  number={3},
  pages={329},
  year={1991},
  publisher={American Psychological Association}
}

@article{Rapid,
  title={Rapid: Efficient retrieval-augmented long text generation with writing planning and information discovery},
  author={Gu, Hongchao and Li, Dexun and Dong, Kuicai and Zhang, Hao and Lv, Hang and Wang, Hao and Lian, Defu and Liu, Yong and Chen, Enhong},
  journal={arXiv preprint arXiv:2503.00751},
  year={2025}
}

@inproceedings{progressive_disclousure,
  title={Progressive disclosure: empirically motivated approaches to designing effective transparency},
  author={Springer, Aaron and Whittaker, Steve},
  booktitle={Proceedings of the 24th international conference on intelligent user interfaces},
  pages={107--120},
  year={2019}
}

@article{context1,
  title={Learning from texts: Effects of prior knowledge and text coherence},
  author={McNamara, Danielle S and Kintsch, Walter},
  journal={Discourse processes},
  volume={22},
  number={3},
  pages={247--288},
  year={1996},
  publisher={Taylor \& Francis}
}

@article{context2,
  title={Deep-level comprehension of science texts: The role of the reader and the text},
  author={Best, Rachel M and Rowe, Michael and Ozuru, Yasuhiro and McNamara, Danielle S},
  journal={Topics in language disorders},
  volume={25},
  number={1},
  pages={65--83},
  year={2005},
  publisher={LWW}
}

@article{DEER,
  title={DEER: Descriptive knowledge graph for explaining entity relationships},
  author={Huang, Jie and Zhu, Kerui and Chang, Kevin Chen-Chuan and Xiong, Jinjun and Hwu, Wen-mei},
  journal={arXiv preprint arXiv:2205.10479},
  year={2022}
}

@article{LokahiPrototype,
  title={The lokahi prototype: Toward the automatic extraction of entity relationship models from text},
  author={Kaufmann, Michael},
  journal={arXiv preprint arXiv:2201.05327},
  year={2022}
}

@inproceedings{textBaseline1,
  title={Eye tracking on text reading with visual enhancements},
  author={Huth, Franziska and Koch, Maurice and Awad-Mohammed, Miriam and Weiskopf, Daniel and Kurzhals, Kuno},
  booktitle={Proceedings of the 2024 Symposium on Eye Tracking Research and Applications},
  pages={1--7},
  year={2024}
}

@inproceedings{textBaseline2,
  title={An empirical study on the efficiency of graphical vs. textual representations in requirements comprehension},
  author={Sharafi, Zohreh and Marchetto, Alessandro and Susi, Angelo and Antoniol, Giuliano and Gu{\'e}h{\'e}neuc, Yann-Ga{\"e}l},
  booktitle={2013 21st International Conference on Program Comprehension (ICPC)},
  pages={33--42},
  year={2013},
  organization={IEEE}
}



\end{document}